\documentclass[12pt]{iopart}

\newcommand{\dd}{\,\mathrm{d}}

\usepackage{bm}
\usepackage{iopams}
\usepackage{braket}
\usepackage{setstack}
\usepackage{citesort}
\usepackage{longtable}
\usepackage{multirow}
\usepackage{graphicx}
\usepackage{ntheorem}
\usepackage{stackrel}

\newtheorem*{conjec}{Conjecture}

\begin{document}

\title{Skewness of von Neumann entanglement entropy}
\author{Lu Wei}
\address{Department of Electrical and Computer Engineering, University of Michigan - Dearborn, Michigan 48128, USA}
\ead{luwe@umich.edu}
\vspace{10pt}

\begin{abstract}
We study quantum bipartite systems in a random pure state, where von Neumann entropy is considered as a measure of the entanglement. Expressions of the first and second exact cumulants of von Neumann entropy, relevant respectively to the average and fluctuation behavior, are known in the literature. The focus of this paper is on its skewness that specifies the degree of asymmetry of the distribution. Computing the skewness requires additionally the third cumulant, an exact formula of which is the main result of this work. In proving the main result, we obtain as a byproduct various summation identities involving polygamma and related functions. The derived third cumulant also leads to an improved approximation to the distribution of von Neumann entropy.
\end{abstract}


\vspace{2pc}

\maketitle

\section{Introduction and the main result}
Classical information theory is the theory behind modern development of computing, communications, and other fields. As its classical counterpart, quantum information theory aims at understanding the theoretical underpinnings of quantum science and technology. One of the most fundamental features of quantum mechanics is the phenomenon of entanglement, which is the resource and medium that enable quantum technologies.

In this work, we consider the quantum bipartite model proposed in the seminal work of Page~\cite{Page93} in the year $1993$, which becomes a standard model in describing the interaction of a physical object and its environment. For such a model, we wish to understand the degree of entanglement as measured by the von Neumann entropy, the statistical behavior of which can be studied from its cumulants/moments. In principle, the knowledge of all moments determines uniquely the distribution of von Neumann entropy due to its compact support (a.k.a. Hausdorff's moment problem). In practice, a finite number of cumulants can be utilized to construct approximations to the distribution of the entropy, where the higher cumulants describe the tail distribution. The higher cumulants also provide information such as whether the average entropy is typical~\cite{Bianchi19}. In the literature, the mean and variance of von Neumann entropy have been investigated in~\cite{Page93,Foong94,Ruiz95,VPO16,Wei17,Bianchi19} among others. The focus of this paper is to study the skewness (involves the second and third cumulants) of von Neumann entropy that measures the degree of asymmetry of the distribution.

The bipartite model proposed by Page~\cite{Page93} is formulated as follows. Consider a composite quantum system that consists of two subsystems $A$ and $B$ of Hilbert space dimensions $m$ and $n$, respectively. The Hilbert space $\mathcal{H}_{A+B}$ of the composite system is given by the tensor product of the Hilbert spaces of the subsystems, $\mathcal{H}_{A+B}=\mathcal{H}_{A}\otimes\mathcal{H}_{B}$. A random pure state of the composite system is written as a linear combination of the random coefficients $x_{i,j}$ and the complete basis $\left\{\Ket{i^{A}}\right\}$ and $\left\{\Ket{j^{B}}\right\}$ of $\mathcal{H}_{A}$ and $\mathcal{H}_{B}$,
\begin{equation}
\Ket{\psi}=\sum_{i=1}^{m}\sum_{j=1}^{n}x_{i,j}\Ket{i^{A}}\otimes\Ket{j^{B}}.
\end{equation}
The corresponding density matrix in the random pure state is
\begin{equation}\label{eq:rho}
\rho=\Ket{\psi}\Bra{\psi}=\sum_{i,k=1}^{m}\sum_{j,l=1}^{n}x_{i,j}x_{k,l}^{\dag}\Ket{i^{A}}\Bra{k^{A}}\otimes\Ket{j^{B}}\Bra{l^{B}},
\end{equation}
which has the natural constraint $\tr(\rho)=1$ (or equivalently $\braket{\psi|\psi}=1$). This implies that the $m\times n$ random coefficient matrix $\mathbf{X}=(x_{i,j})$ satisfies
\begin{equation}\label{eq:pcv}
\tr\left(\mathbf{XX}^{\dag}\right)=1.
\end{equation}
We assume without loss of generality that $m\leq n$. The reduced density matrix $\rho_{A}$ of the smaller subsystem $A$ is computed by partial tracing of the full density matrix~(\ref{eq:rho}) over the other subsystem $B$ (interpreted as the environment) as
\begin{equation}
\rho_{A}=\tr_{B}(\rho)=\sum_{i,k=1}^{m}\sum_{j=1}^{n}x_{i,j}x_{k,j}^{\dag}\Ket{i^{A}}\Bra{k^{A}}=\sum_{i,k=1}^{m}w_{i,k}\Ket{i^{A}}\Bra{k^{A}},
\end{equation}
where $w_{i,k}$ is the $(i,k)$-th entry of the $m\times m$ Hermitian matrix $\mathbf{W}=\mathbf{XX}^{\dag}$. The Schmidt decomposition of $\rho_{A}$ is given by
\begin{equation}
\rho_{A}=\sum_{i=1}^{m}\lambda_{i}\Ket{\phi_{i}^{A}}\Bra{\phi_{i}^{A}},
\end{equation}
where $0<\lambda_{m}<\dots<\lambda_{1}<1$ are the eigenvalues of $\mathbf{W}$, and the condition~(\ref{eq:pcv}) now implies the fixed-trace constraint
\begin{equation}\label{eq:lamcon}
\sum_{i=1}^{m}\lambda_{i}=1.
\end{equation}
The probability measure of $\rho_{A}$ is the Haar measure satisfying the additional constraint~(\ref{eq:lamcon}). The corresponding eigenvalue density of $\mathbf{W}$ is well-known (see, e.g.,~\cite{Page93})
\begin{equation}\label{eq:fte}
f\left(\bm{\lambda}\right)=\frac{\Gamma(mn)}{C}~\delta\left(1-\sum_{i=1}^{m}\lambda_{i}\right)\prod_{1\leq i<j\leq m}\left(\lambda_{i}-\lambda_{j}\right)^{2}\prod_{i=1}^{m}\lambda_{i}^{n-m},
\end{equation}
where $\delta(\cdot)$ is the Dirac delta function and the constant
\begin{equation}\label{eq:con}
C=\prod_{i=1}^{m}\Gamma(n-i+1)\Gamma(i).
\end{equation}
The random matrix ensemble~(\ref{eq:fte}) is also known as the (unitary) fixed-trace ensemble. The above discussed bipartite model is useful in describing the interaction of various real-world quantum systems. For example, in~\cite{Page93} the subsystem $A$ is the black hole and the subsystem $B$ is the associated radiation field. In another example~\cite{Majumdar}, the subsystem $A$ is a set of spins and the subsystem $B$ represents the environment of a heat bath.

The degree of entanglement of subsystems can be measured by entanglement entropies, which are functions of eigenvalues of $\mathbf{W}$. We consider the standard measure of von Neumann entropy of the subsystem\footnote{Note that since the composite system is in a random pure state, the von Neumann entropy of the full system is zero~\cite{Page93}.}
\begin{equation}\label{eq:vN}
S=-\tr\left(\rho_{A}\ln\rho_{A}\right)=-\sum_{i=1}^{m}\lambda_{i}\ln\lambda_{i},~~~~~~S\in\left[0, \ln{m}\right],
\end{equation}
which achieves the separable state ($S=0$) when $\lambda_{1}=1$, $\lambda_{2}=\dots=\lambda_{m}=0$ and the maximally-entangled state ($S=\ln{m}$) when $\lambda_{1}=\dots\lambda_{m}=1/m$. Statistical information of the von Neumann entropy is encoded through its moments
\begin{equation}
\mathbb{E}_{f}\!\left[S^{k}\right],~~~~~~k=1,2,3,\dots,
\end{equation}
where the expectation is taken over the density~(\ref{eq:fte}). In general, the moment sequence, $m_{1},m_{2},m_{3},\dots$, and the cumulant sequence, $\kappa_{1},\kappa_{2},\kappa_{3},\dots$, for a random variable are related: the $i$-th moment is an $i$-th degree polynomial in the first $i$ cumulants and vice versa. In particular, the relation pairs up to $i=3$ are
\begin{eqnarray}
m_{1} &= \kappa_{1} \label{eq:1km} \\
m_{2} &= \kappa_{2}+\kappa_{1}^{2}~~~~~~~~~~~&\kappa_{2} = m_{2}-m_{1}^{2} \label{eq:2mk} \\
m_{3} &= \kappa_{3}+3\kappa_{2}\kappa_{1}+\kappa^{3}_{1}~~~~~~~~~~~&\kappa_{3} = m_{3}-3m_{2}m_{1}+2m_{1}^{3}. \label{eq:3mk}
\end{eqnarray}
It turns out that the cumulants/moments of von Neumann entropy can be expressed through polygamma functions, the $i$-th order of which is defined as
\begin{equation}\label{eq:pgd}
\psi_{i}(z)=\frac{\partial^{i+1}\ln\Gamma(z)}{\partial z^{i+1}}=(-1)^{i+1}i!\sum_{k=0}^{\infty}\frac{1}{(k+z)^{i+1}}.
\end{equation}
For positive integer arguments, the digamma function ($0$-th order polygamma function) is simplified to a finite sum as
\begin{equation}\label{eq:p0}
\psi_{0}(l)=-\gamma+\sum_{k=1}^{l-1}\frac{1}{k}
\end{equation}
with $\gamma\approx0.5772$ being the Euler's constant, and the polygamma functions of order $j\geq1$ can be also reduced to finite sums as
\begin{equation}\label{eq:pg}
\psi_{j}(l)=(-1)^{j+1}j!\left(\zeta(j+1)-\sum_{k=1}^{l-1}\frac{1}{k^{j+1}}\right),
\end{equation}
where
\begin{equation}\label{eq:z}
\zeta(s)=\sum_{k=1}^{\infty}\frac{1}{k^{s}}
\end{equation}
is the Riemann zeta function. In particular, the present paper involves finite sum form
\begin{equation}\label{eq:p1}
\psi_{1}(l)=\frac{\pi^{2}}{6}-\sum_{k=1}^{l-1}\frac{1}{k^{2}}
\end{equation}
of the trigamma function, and finite sum form
\begin{equation}\label{eq:p2}
\psi_{2}(l)=-2\zeta(3)+2\sum_{k=1}^{l-1}\frac{1}{k^{3}}
\end{equation}
of the second order polygamma function with
\begin{equation}
\zeta(3)\approx1.20206
\end{equation}
being the Ap\'{e}ry's constant.

With the above definitions, we now discuss the state of the art in discovering the exact cumulants of von Neumann entropy. The mean value of von Neumann entropy (first cumulant) relevant to the typical behavior of entanglement was conjectured by Page, in the same work~\cite{Page93} the bipartite model was proposed, as
\begin{equation}\label{eq:k1}
\kappa_{1}=\psi_{0}(mn+1)-\psi_{0}(n)-\frac{m+1}{2n}.
\end{equation}
Page's conjecture was proved shortly afterwards in~\cite{Foong94,Ruiz95} among others. The variance of von Neumann entropy (second cumulant) that describes the fluctuation of entanglement around the typical value was conjectured in~\cite{VPO16} as
\begin{equation}\label{eq:k2}
\kappa_{2}=-\psi_{1}\left(mn+1\right)+\frac{m+n}{mn+1}\psi_{1}\left(n\right)-\frac{(m+1)(m+2n+1)}{4n^{2}(mn+1)}.
\end{equation}
This variance formula was firstly proved in~\cite{Wei17}, and was independently proved in~\cite{Bianchi19} recently, see also~\cite{Wei19} for a discussion on the latter proof. In the present work, we focus on the skewness of von Neumann entropy defined as the third standardized moment
\begin{equation}\label{eq:skew}
\gamma_{1}=\mathbb{E}_{f}\!\left[\left(\frac{S-\kappa_{1}}{\sqrt{\kappa_{2}}}\right)^{3}\right]=\frac{\kappa_{3}}{\kappa_{2}^{3/2}},
\end{equation}
where the notation $\gamma_{1}$ is due to Karl Pearson. The skewness quantifies the (lack of) symmetry of a probability distribution, where a symmetric distribution such as the Gaussian distribution has a skewness of zero. As seen from the definition~(\ref{eq:skew}), calculating the skewness requires the additional knowledge on the third cumulant of $S$, an expression of which is given by
\begin{eqnarray}\label{eq:k3}
\fl\kappa_{3}&\!\!\!\!\!\!\!\!\!\!\!\!\!\!\!\!\!\!\!\!\!\!\!\!\!\!\!\!\!\!=&\!\!\!\!\!\!\!\!\!\!\!\!\!\!\!\!\!\!\!\!\!\!\psi_{2}\left(mn+1\right)-\frac{m^{2}+3mn+n^{2}+1}{(mn+1)(mn+2)}\psi_{2}(n)+\frac{(m^2-1)(mn-3n^2+1)}{n(mn+1)^{2}(mn+2)}\psi_{1}(n+1)-\nonumber\\
&&\!\!\!\!\!\!\!\!\!\!\!\!\!\!\!\!\!\!\!\!\!\!\frac{(m+1)\left(2m^{3}n+3m^{2}n^{2}+2m^{2}+4mn^{3}+15mn^{2}+12mn-2n^{2}+6n+6\right)}{4n^{3}(mn+1)^{2}(mn+2)}.
\end{eqnarray}
Proving the above formula of the exact third cumulant of von Neumann entropy is the main contribution of this work. Note that the third moment formula~(\ref{eq:k3}) has been recently reported in~\cite{Bianchi19}, where the authors described the computations that led to the result~(\ref{eq:k3}) as ``The computation is an herculean task but with the help of Wolfram's Mathematica we are able to simplify the exact formula for $\mu_{3}$''. The rest of the paper is to decode the above description by providing a proof to the claimed result~(\ref{eq:k3}). Note also that despite having a different starting point of the calculation in~\cite[Eq.~(S32)]{Bianchi19} than that in the current paper~(\ref{eq:mST}), the subsequent bulk of the calculations omitted in~\cite{Bianchi19} necessarily involves the calculations performed here in Sections~\ref{sec:IABC} and~\ref{sec:simp}. To see the above statement in the second cumulant computation, we refer to~\cite{Wei19}. Interestingly, as will be seen the proof relies crucially on new summation identities derived in this work that in fact have not been implemented in Mathematica~\cite{KG}.

Approximations to the distribution of von Neumann entropy can be constructed from the closed-form cumulant expressions. For convenience, we first standardize the von Neumann entropy as
\begin{equation}\label{eq:X}
X=\frac{S-\kappa_{1}}{\sqrt{\kappa_{2}}}
\end{equation}
so that the first two cumulants of the random variable $X$ become
\begin{equation}\label{eq:cX12}
\kappa_{1}^{(X)}=0,~~~~~~\kappa_{2}^{(X)}=1.
\end{equation}
The higher order cumulants of $S$ and $X$, beyond the first two in~(\ref{eq:k1}),~(\ref{eq:k2}),~(\ref{eq:cX12}), are related by
\begin{equation}\label{eq:SXc}
\kappa_{j}^{(X)}=\frac{\kappa_{j}}{\kappa_{2}^{j/2}},~~~~~~j\geq3.
\end{equation}
In principle, the probability density function of the standardized variable $X$ can be represented as~\cite{Cramer}
\begin{equation}\label{eq:app}
f_{X}(x)=\varphi_{X}(x)+r_{X}(x),
\end{equation}
where the function $r_{X}(x)$ is the reminder term of the initial approximation $\varphi_{X}(x)$. Since the variable~(\ref{eq:X}) is supported in $X\in(-\infty,\infty)$ with the first two cumulants given by~(\ref{eq:cX12}), we consider a standard Gaussian distribution as the initial approximation, i.e.,
\begin{equation}\label{eq:iappr}
f_{X}(x)\approx\varphi_{X}(x)=\frac{1}{\sqrt{2\pi}}\e^{-\frac{1}{2}x^{2}}.
\end{equation}
The corresponding initial Gaussian approximation to $S$ incorporates its first two cumulants~(\ref{eq:k1}) and~(\ref{eq:k2}) through the affine transformation~(\ref{eq:X}). The reminder term $r_{X}(x)$ associated with $\varphi_{X}(x)$ in~(\ref{eq:iappr}) admits different types of expansions. We shall adopt an expansion based on orthogonal polynomials that are induced from the initial approximation $\varphi_{X}(x)$, where the reminder term is formally expanded as~\cite{Cramer}
\begin{equation}\label{eq:expan}
r_{X}(x)=\sum_{k=1}^{\infty}\frac{d_{k}}{k!}\varphi_{X}^{(k)}(x).
\end{equation}
The $k$-th derivative $\varphi_{X}^{(k)}(x)$ of the Gaussian distribution~(\ref{eq:iappr}) gives rise to the (probabilists') Hermite polynomials $H_{k}(x)$ of degree $k$ as
\begin{equation}\label{eq:reH}
\varphi_{X}^{(k)}(x)=(-1)^{k}H_{k}(x)\varphi_{X}(x).
\end{equation}
The Hermite polynomials satisfy the orthogonality relation~\cite{Cramer,Forrester}
\begin{equation}\label{eq:ocH}
\int_{-\infty}^{\infty}\!\!\varphi_{X}(x)H_{k}(x)H_{l}(x)\dd{x}=k!\delta_{kl}
\end{equation}
with $\delta_{kl}$ being the Kronecker delta function, where the first a few of $H_{k}(x)$ are
\begin{eqnarray}
H_{0}(x) &=1,~~~~~~~~~ &H_{1}(x)=x \label{eq:H01} \\
H_{2}(x) &=x^{2}-1,~~~~~~~~~ &H_{3}(x)=x^{3}-3x. \label{eq:H23}
\end{eqnarray}
With the choices of $\varphi_{X}(x)$ in~(\ref{eq:iappr}) and $r_{X}(x)$ in~(\ref{eq:expan}), the expansion~(\ref{eq:app}) is also known as the type-A Gram-Charlier series~\cite{Cramer}. Its $k$-th coefficient $d_{k}$ can be conveniently expressed as a polynomial in the first $k$ cumulants of the standardized random variable $X$ in~(\ref{eq:X}). In particular, it can be directly verified by the orthogonality relation~(\ref{eq:ocH}) and the results~(\ref{eq:H01}),~(\ref{eq:H23}) that
\begin{equation}
d_{1}=d_{2}=0
\end{equation}
and that $d_{3}$ equals the negative of $\kappa_{3}^{(X)}$, which also equals the negative of the skewness of $S$ (cf.~(\ref{eq:SXc}) and~(\ref{eq:skew})), as
\begin{equation}
d_{3}=-\kappa_{3}^{(X)}=-\frac{\kappa_{3}}{\kappa_{2}^{3/2}}=-\gamma_{1}.
\end{equation}
As a result, a refined approximation (cf.~(\ref{eq:iappr})) to the distribution of standardized von Neumann entropy~(\ref{eq:X}) is obtained as
\begin{eqnarray}
f_{X}(x)&=&\varphi_{X}(x)+\sum_{k=1}^{\infty}\frac{d_{k}}{k!}\varphi_{X}^{(k)}(x)\\
&\approx&\varphi_{X}(x)+\frac{d_{3}}{3!}\varphi_{X}^{(3)}(x), \label{eq:k3app}
\end{eqnarray}
where the correction term
\begin{equation}\label{eq:ct}
\frac{d_{3}}{3!}\varphi_{X}^{(3)}(x)=\frac{\kappa_{3}}{6\kappa_{2}^{3/2}}H_{3}(x)\varphi_{X}(x)
\end{equation}
incorporates the derived third cumulant~(\ref{eq:k3}).
\begin{figure}[!h]
\centering
\includegraphics[width=0.92\linewidth]{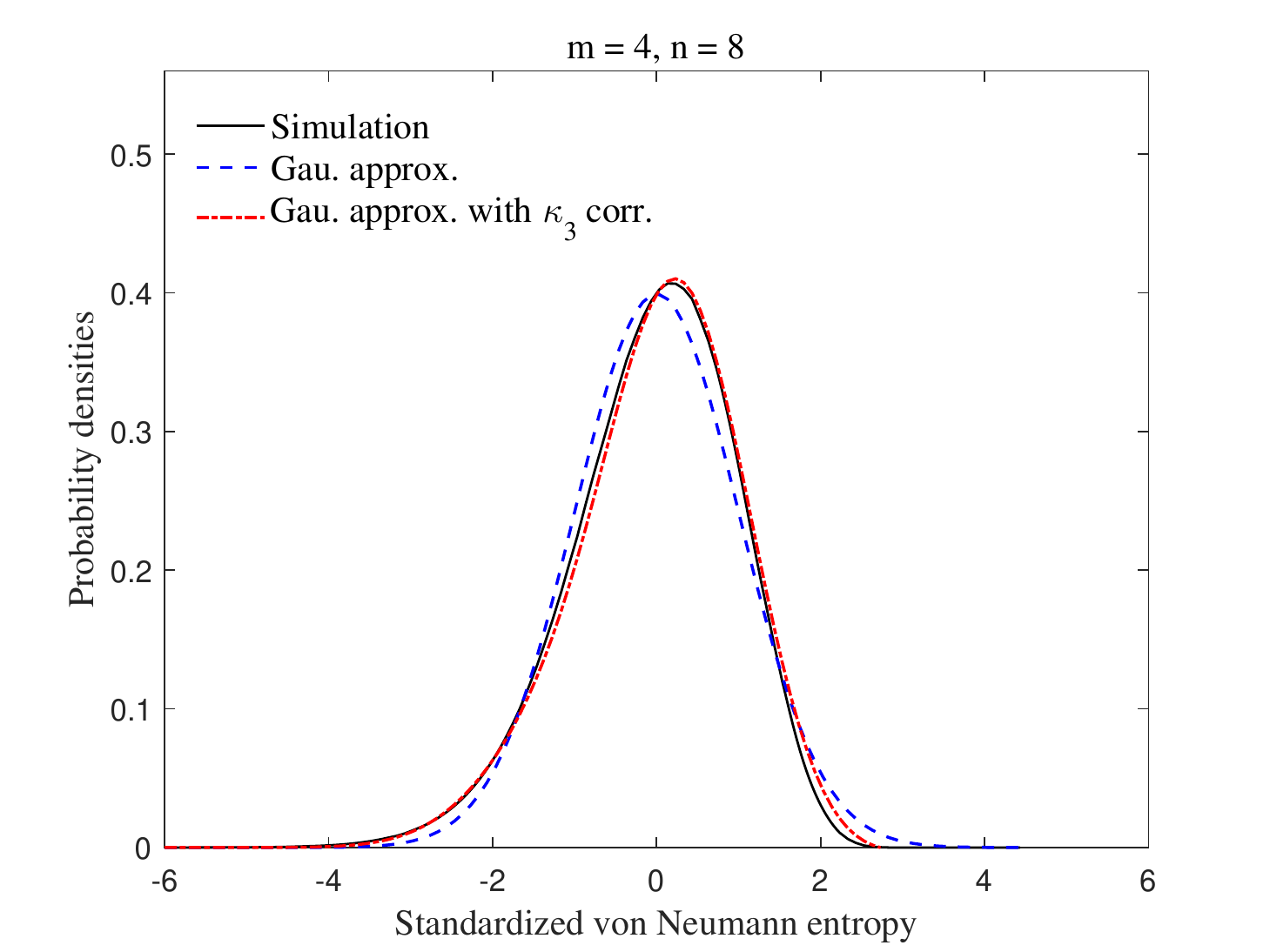}
\caption{Probability densities of standardized von Neumann entropy~(\ref{eq:X}) of subsystem dimensions $m=4$ and $n=8$: A comparison of Gaussian approximation~(\ref{eq:iappr}) (dashed line in blue) to the refined approximation~(\ref{eq:k3app}) (dash-dot line in red) with the $\kappa_{3}$ correction term~(\ref{eq:ct}). The solid line in black represents simulated true distribution.}
\label{fig:p1}
\end{figure}
In Figure~\ref{fig:p1}, we numerically compare the refined approximation~(\ref{eq:k3app}) to the initial Gaussian approximation~(\ref{eq:iappr}), where the dimensions of the subsystems are $m=4$ and $n=8$. Comparing with the simulated true distribution, it is observed that the new approximation~(\ref{eq:k3app}) that incorporates the additional knowledge on the skewness via the correction term~(\ref{eq:ct}) is more accurate than the first two cumulants based Gaussian approximation~(\ref{eq:iappr}). As compared to the symmetric Gaussian distribution, we also see from Figure~\ref{fig:p1} that the true distribution of von Neumann entropy is indeed a skewed one. The distribution appears to be left-skewed (a.k.a. a negative skewness), where the left tail of the distribution is longer.
\begin{figure}[!h]
\centering
\includegraphics[width=0.92\linewidth]{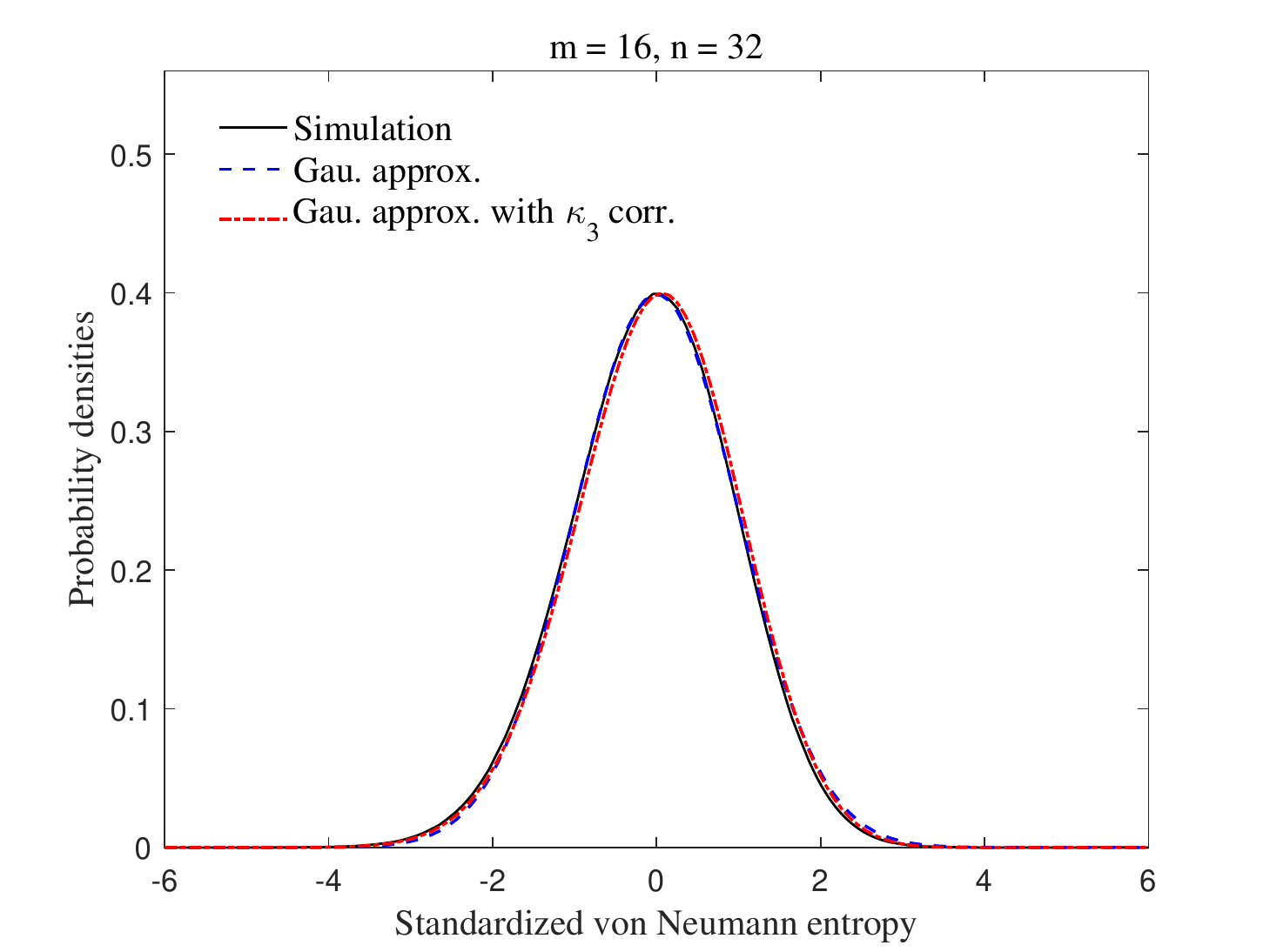}
\caption{Probability densities of standardized von Neumann entropy~(\ref{eq:X}) of subsystem dimensions $m=16$ and $n=32$: A numerical support to the conjectured asymptotic limit. The solid line in black, the dashed line in blue, and the dash-dot line in red represents the simulated true distribution, the standard Gaussian distribution~(\ref{eq:iappr}), and the refined approximation~(\ref{eq:k3app}), respectively.}
\label{fig:p2}
\end{figure}
In Figure~\ref{fig:p2}, we perform another numerical study by simultaneously increasing the subsystem dimensions to $m=16$ and $n=32$ with their ratio
\begin{equation}
c=\frac{m}{n}=\frac{1}{2}
\end{equation}
kept the same as the value in Figure~\ref{fig:p1}. We observe from Figure~\ref{fig:p2} that both the simulated and the approximate~(\ref{eq:k3app}) distributions of $X$ approach the standard Gaussian distribution~(\ref{eq:iappr}), where the three curves almost overlap. As the author has learnt from Sean O'Rourke, the observed asymptotic Gaussian behavior is typical for a wide class of linear spectral statistics\footnote{A linear spectral statistics can be defined as $\sum_{k=1}^{m}f(x_{i})$, where $x_{i}, i=1,\dots,m$, are the eigenvalues of an $m\times m$ Hermitian random matrix.} over different random matrix ensembles. In fact, we conjecture the following central limit theorem for the linear spectral statistics of von Neumann entropy~(\ref{eq:vN}) over the fixed-trace ensemble~(\ref{eq:fte}).
\begin{conjec}[O'Rourke-Wei]
In the limit
\begin{equation}\label{eq:lim}
m\to\infty,~~~~n\to\infty,~~~~\frac{m}{n}=c\in(0,1],
\end{equation}
the standardized von Neumann entropy defined in~(\ref{eq:X}) converges in distribution to a Gaussian random variable with zero mean and unit variance.
\end{conjec}
Note that the regime~(\ref{eq:lim}) is a typical asymptotic regime in random matrix theory (a.k.a. high-dimensional asymptotic regime), where both dimensions $m$ and $n$ approach infinity. This is different from the classical asymptotic regime as studied in, e.g.,~\cite{Page93,Bianchi19}, where the dimension $m$ is fixed as $n$ goes to infinity. Proving the above conjecture requires showing that all the higher cumulants~(\ref{eq:SXc}) vanish in the regime~(\ref{eq:lim}), i.e.,
\begin{equation}\label{eq:ks}
\kappa_{j}^{(X)}=\frac{\kappa_{j}}{\kappa_{2}^{j/2}}\rightarrow0,~~~~~~j\geq3.
\end{equation}
With the obtained main result~(\ref{eq:k3}) of this work, we can show that the third cumulant of $X$ vanishes in the regime~(\ref{eq:lim}) as follows. First, by the asymptotic behavior of polygamma functions
\begin{equation}
\psi_{j}(x)=\Theta\left(\frac{1}{x^{j}}\right),~~~~~~x\to\infty,~~~~~~j\geq1,
\end{equation}
the second cumulant~(\ref{eq:k2}) and third cumulant~(\ref{eq:k3}) of $S$ scale in the limit~(\ref{eq:lim}) as
\begin{equation}\label{eq:k2a}
\kappa_{2}=\Theta\left(\frac{1}{n^{2}}\right),
\end{equation}
and
\begin{equation}\label{eq:k3a}
\kappa_{3}=\Theta\left(\frac{1}{n^{4}}\right),
\end{equation}
respectively. Consequently, as claimed, the third cumulant of $X$ (or the skewness of $S$)
\begin{equation}
\kappa_{3}^{(X)}=\frac{\kappa_{3}}{\kappa_{2}^{3/2}}=\frac{\Theta(1/n^{4})}{\Theta(1/n^{3})}=\Theta\left(\frac{1}{n}\right)
\end{equation}
vanishes in the regime~(\ref{eq:lim}). In addition, based on the structure of the first three exact cumulants and the asymptotic results~(\ref{eq:k2a}),~(\ref{eq:k3a}), we further conjecture the limiting behavior of all higher cumulants as
\begin{equation}
\kappa_{j}=\Theta\left(\frac{1}{n^{2j-2}}\right),~~~~~~j\geq4,
\end{equation}
or equivalently, cf.~(\ref{eq:ks}) and~(\ref{eq:k2a}),
\begin{equation}
\kappa_{j}^{(X)}=\Theta\left(\frac{1}{n^{j-2}}\right),~~~~~~j\geq4.
\end{equation}

The rest of the paper is organized as follows. In Section~\ref{sec:deri}, we derive the main result~(\ref{eq:k3}) on the third cumulant of von Neumann entropy. Specifically, in Section~\ref{sec:rela} the original problem is reduced to the task of computing three integrals~(\ref{eq:IA}),~(\ref{eq:IB}), and~(\ref{eq:IC}) by exploring the relation to a more convenient random matrix ensemble. The three integrals are calculated explicitly in Section~\ref{sec:IABC} to the form of finite summations involving polygamma functions. The remaining part of the derivation is performed in Section~\ref{sec:simp}, where we evaluate and simplify the resulting sums with the help of two types of polygamma summation identities. The relevant summation identities are listed in the appendices, where we also discuss the strategies in finding them.

\section{Derivation of the third cumulant}\label{sec:deri}

\subsection{Cumulant relation}\label{sec:rela}
The task of this subsection is to convert the third cumulant of von Neumann entropy to that of a related random variable, the computation of which can then be conveniently performed.

We start by studying the relation between the third moments. By construction, the random coefficient matrix $\mathbf{X}$ has a natural relation with a Wishart matrix $\mathbf{YY}^{\dag}$ as
\begin{equation}\label{eq:wf}
\mathbf{XX}^{\dag}=\frac{\mathbf{YY}^{\dag}}{\tr\left(\mathbf{YY}^{\dag}\right)},
\end{equation}
where $\mathbf{Y}$ is an $m\times n$ ($m\leq n$) matrix of independently and identically distributed complex Gaussian entries. The density of the eigenvalues $0<\theta_{m}<\dots<\theta_{1}<\infty$ of $\mathbf{YY}^{\dag}$ equals~\cite{Forrester}
\begin{equation}\label{eq:we}
g\left(\bm{\theta}\right)=\frac{1}{C}\prod_{1\leq i<j\leq m}\left(\theta_{i}-\theta_{j}\right)^{2}\prod_{i=1}^{m}\theta_{i}^{n-m}\e^{-\theta_{i}},
\end{equation}
where $C$ is given by~(\ref{eq:con}) and the above ensemble is known as the Wishart-Laguerre ensemble. The trace of the Wishart matrix
\begin{equation}
r=\tr\left(\mathbf{YY}^{\dag}\right)=\sum_{i=1}^{m}\theta_{i}
\end{equation}
follows a gamma distribution with the density
\begin{equation}\label{eq:r}
h_{mn}(r)=\frac{1}{\Gamma(mn)}\e^{-r}r^{mn-1},~~~~~~r\in[0,\infty).
\end{equation}
The relation~(\ref{eq:wf}) induces the change of variables
\begin{equation}\label{eq:cv}
\lambda_{i}=\frac{\theta_{i}}{r},~~~~~~i=1,\ldots,m,
\end{equation}
that leads to a well-known relation (see, e.g.~\cite{Page93}) among the densities~(\ref{eq:fte}),~(\ref{eq:we}), and~(\ref{eq:r}) as
\begin{equation}\label{eq:relation}
f\left(\bm{\lambda}\right)h_{mn}(r)\dd r\prod_{i=1}^{m}\dd\lambda_{i}=g\left(\bm{\theta}\right)\prod_{i=1}^{m}\dd\theta_{i}.
\end{equation}
The above relation implies that $r$ is independent of each $\lambda_{i}$, $i=1,\ldots,m$, since their densities factorize. We now define the random variable
\begin{equation}\label{eq:T}
T=\sum_{i=1}^{m}\theta_{i}\ln\theta_{i},
\end{equation}
as the induced entropy\footnote{For convenience of the discussion, we refer to the random variable $T$ as an induced entropy, which may not have the physical meaning of an entropy.} over the Wishart-Laguerre ensemble~(\ref{eq:we}). The relation~(\ref{eq:relation}) has been utilized to convert the first and second moment of $S$ to $T$ in~\cite{Page93} and~\cite{Wei17}, respectively. In a similar manner, the computation of the third moment of $S$ can be also converted to that of $T$ as follows. First, by the change of variables~(\ref{eq:cv}), one has
\begin{equation}\label{eq:ST}
S=-\sum_{i=1}^{m}\frac{\theta_{i}}{r}\ln\frac{\theta_{i}}{r}=r^{-1}\left(r\ln r-T\right),
\end{equation}
and consequently
\begin{eqnarray}
S^{3} &=& r^{-3}\left(-T^{3}+T^{2}3r\ln r-T3r^{2}\ln^{2}r+r^{3}\ln^{3}r\right)\\
&=& r^{-3}\left(-T^{3}+S^{2}3r^{3}\ln r-S3r^{3}\ln^{2}r+r^{3}\ln^{3}r\right),\label{eq:TS}
\end{eqnarray}
where the second equality is obtained by replacing $T$ (except for the the highest power term $T^{3}$) by $S$ using the relation~(\ref{eq:ST}). As will be seen, the form (\ref{eq:TS}) makes it possible to utilize the independence between $r$ and $\bm{\lambda}$ so as to perform the subsequent calculations. The third moment of $S$ can then be computed as
\begin{eqnarray}
\fl\mathbb{E}_{f}\!\left[S^{3}\right]&\!\!\!\!\!\!\!\!\!\!\!\!\!\!\!\!\!=&\int_{\bm{\lambda}}r^{-3}\left(-T^{3}+S^{2}3r^{3}\ln r-S3r^{3}\ln^{2}r+r^{3}\ln^{3}r\right)f\left(\bm{\lambda}\right)\prod_{i=1}^{m}\dd\lambda_{i} \label{eq:S2T} \\
&\!\!\!\!\!\!\!\!\!\!\!\!\!\!\!\!\!=&\int_{\bm{\lambda}}r^{-3}\left(-T^{3}+S^{2}3r^{3}\ln r-S3r^{3}\ln^{2}r+r^{3}\ln^{3}r\right)f\left(\bm{\lambda}\right)\prod_{i=1}^{m}\dd\lambda_{i}\int_{r}h_{mn+3}(r)\dd{r} \nonumber\\
&\!\!\!\!\!\!\!\!\!\!\!\!\!\!\!\!\!=&\frac{-\mathbb{E}_{g}\!\left[T^{3}\right]+3\mathbb{E}_{h}\!\left[r^{3}\ln r\right]\mathbb{E}_{f}\!\left[S^{2}\right]-3\mathbb{E}_{h}\!\left[r^{3}\ln^{2}r\right]\mathbb{E}_{f}\!\left[S\right]+\mathbb{E}_{h}\!\left[r^{3}\ln^{3}r\right]}{(mn)_{3}}\label{eq:mST},
\end{eqnarray}
where in the second line we multiple an appropriate constant
\begin{equation}
1=\int_{r}h_{mn+3}(r)\dd{r},
\end{equation}
and the last equality~(\ref{eq:mST}) is obtained by the using identity (with $(a)_{n}$ denoting the Pochhammer's symbol)
\begin{equation}
r^{-3}h_{mn+3}(r)=\frac{h_{mn}(r)}{(mn)_{3}},
\end{equation}
as well as the change of measures~(\ref{eq:relation}) for the $T^{3}$ term along with the fact that $r$ and $\bm{\lambda}$ are independent (so that the integrals involving $r$ and $S$ are evaluated separately). In~(\ref{eq:mST}), the expected values over the density $h_{mn}(r)$ in~(\ref{eq:r}) are computed by the identities
\begin{eqnarray}
\int_{0}^{\infty}\!\e^{-r}r^{a-1}\ln{r}\dd{r}&=&\Gamma(a)\psi_{0}(a) \\
\int_{0}^{\infty}\!\e^{-r}r^{a-1}\ln^{2}{r}\dd{r}&=&\Gamma(a)\left(\psi_{0}^{2}(a)+\psi_{1}(a)\right) \\
\int_{0}^{\infty}\!\e^{-r}r^{a-1}\ln^{3}{r}\dd{r}&=&\Gamma(a)\left(\psi_{0}^{3}(a)+3\psi_{0}(a)\psi_{1}(a)+\psi_{2}(a)\right),
\end{eqnarray}
obtained by taking derivatives with respect to the parameter $a$ of gamma function
\begin{equation}
\int_{0}^{\infty}\!\e^{-r}r^{a-1}\dd{r}=\Gamma(a),
\end{equation}
as
\begin{eqnarray}
\fl\mathbb{E}_{h}\!\left[r^{3}\ln r\right] &=& (mn)_{3}\psi_{0}(mn+3) \\
\fl\mathbb{E}_{h}\!\left[r^{3}\ln^{2}r\right] &=& (mn)_{3}\left(\psi_{0}^{2}(mn+3)+\psi_{1}(mn+3)\right) \\
\fl\mathbb{E}_{h}\!\left[r^{3}\ln^{3}r\right] &=& (mn)_{3}\left(\psi_{0}^{3}(mn+3)+3\psi_{0}(mn+3)\psi_{1}(mn+3)+\psi_{2}(mn+3)\right).
\end{eqnarray}
The first and second moment of $S$ in~(\ref{eq:mST}) are also known, cf.~(\ref{eq:1km}),~(\ref{eq:2mk}),~(\ref{eq:k1}), and~(\ref{eq:k2}),
\begin{eqnarray}
\mathbb{E}_{f}\!\left[S\right] &=& \kappa_{1} \label{eq:S1} \\
\mathbb{E}_{f}\!\left[S^{2}\right] &=& \kappa_{2}+\kappa_{1}^{2}. \label{eq:S2}
\end{eqnarray}
Therefore, to obtain $\mathbb{E}_{f}\!\left[S^{3}\right]$ the remaining term to compute in~(\ref{eq:mST}) is $\mathbb{E}_{g}\!\left[T^{3}\right]$. Since
\begin{equation*}
\!\!\!\!\!\!\!\!\!\!\!\!\!\!\!\!\!\!\!\!\!\!\!\!\!\!\!T^{3}=\sum_{i=1}^{m}\theta_{i}^{3}\ln^{3}\theta_{i}+3\sum_{1\leq i\neq j\leq m}\theta_{i}^{2}\theta_{j}\ln^{2}\theta_{i}\ln\theta_{j}+6\sum_{1\leq i\neq j\neq k\leq m}\theta_{i}\theta_{j}\theta_{k}\ln\theta_{i}\ln\theta_{j}\ln\theta_{k},
\end{equation*}
the computation of $\mathbb{E}_{g}\!\left[T^{3}\right]$ involves one, two, and three arbitrary eigenvalue densities, denoted respectively by $g_{1}(x_{1})$, $g_{2}(x_{1},x_{2})$, and $g_{3}(x_{1},x_{2},x_{3})$, of the Wishart-Laguerre ensemble~(\ref{eq:we}) as
\begin{eqnarray*}
\fl\mathbb{E}_{g}\!\left[T^{3}\right] &=& {m\choose1}\int_{0}^{\infty}\!\!x_{1}^{3}\ln^{3}x_{1}~g_{1}(x_{1})\dd x_{1}+3{m\choose1}{m-1\choose1}\int_{0}^{\infty}\!\!\int_{0}^{\infty}\!\!x_{1}^{2}x_{2}\ln^{2}x_{1}\ln x_{2}\times \\
\fl&&\!\!\!\!\!\!\!\!\!\!\!\!\!\!\!\!\!\!\!\!\!\!\!\!\!\!\!g_{2}\left(x_{1},x_{2}\right)\dd x_{1}\dd x_{2}+6{m\choose3}\int_{0}^{\infty}\!\!\int_{0}^{\infty}\!\!\int_{0}^{\infty}\!\!x_{1}x_{2}x_{3}\ln x_{1}\ln x_{2}\ln x_{3}~g_{3}\left(x_{1},x_{2},x_{2}\right)\dd x_{1}\dd x_{2}\dd x_{3}.
\end{eqnarray*}
It is a well-known result in random matrix theory that the joint density $g_{N}(x_{1},\dots,x_{N})$ of $N$ (out of $m$) arbitrary eigenvalues of various matrix models, including the Wishart-Laguerre ensemble, can be written in terms of a determinant of a correlation kernel $K(x_{i},x_{j})$ as~\cite{Forrester}
\begin{equation}\label{eq:Nei}
g_{N}(x_{1},\dots,x_{N})=\frac{(m-N)!}{m!}\det\left(K\left(x_{i},x_{j}\right)\right)_{i,j=1}^{N}.
\end{equation}
The determinant in~(\ref{eq:Nei}) is known as the $N$-point correlation function, where the symmetric correlation kernel $K(x_{i},x_{j})$ uniquely specifies the random matrix ensemble. As a result of~(\ref{eq:Nei}), the arbitrary eigenvalue densities needed to compute $\mathbb{E}_{g}\!\left[T^{3}\right]$ are
\begin{eqnarray}
\fl g_{1}(x_{1}) &\!\!\!\!=& \frac{1}{m}K\left(x_{1},x_{1}\right) \label{eq:g1} \\
\fl g_{2}(x_{1},x_{2}) &\!\!\!\!=& \frac{1}{m(m-1)}\left(K(x_{1},x_{1})K(x_{2},x_{2})-K^{2}(x_{1},x_{2})\right) \\
\fl g_{3}(x_{1},x_{2},x_{3}) &\!\!\!\!=& \frac{1}{m(m-1)(m-2)}(K(x_{1},x_{1})K(x_{2},x_{2})K(x_{3},x_{3})+K(x_{1},x_{2})K(x_{2},x_{3})\times\nonumber\\ &&K(x_{3},x_{1})+K(x_{1},x_{3})K(x_{3},x_{2})K(x_{2},x_{1})-K(x_{1},x_{2})K(x_{2},x_{1})K(x_{3},x_{3})\nonumber \\
&&-K(x_{1},x_{3})K(x_{3},x_{1})K(x_{2},x_{2})-K(x_{2},x_{3})K(x_{3},x_{2})K(x_{1},x_{1})),
\end{eqnarray}
and consequently the third moment of $T$ can be written as
\begin{eqnarray}\label{eq:T3}
\fl\mathbb{E}_{g}\!\left[T^{3}\right]&=&\left(\int_{0}^{\infty}\!\!x\ln{x}~K(x,x)\dd{x}\right)^{3}+3\int_{0}^{\infty}\!\!x\ln{x}~K(x,x)\dd{x}\int_{0}^{\infty}\!\!x^{2}\ln^{2}{x}~K(x,x)\dd{x}+ \nonumber \\
\fl&&\int_{0}^{\infty}\!\!x^{3}\ln^{3}{x}~K(x,x)\dd{x}-3\int_{0}^{\infty}\!\!x\ln{x}~K(x,x)\dd{x}\int_{0}^{\infty}\!\!\int_{0}^{\infty}\!\!xy\ln x\ln y\times\nonumber \\
\fl&&K^{2}(x,y)\dd x\dd y-3\int_{0}^{\infty}\!\!\int_{0}^{\infty}\!\!x^{2}y\ln^{2}x\ln y~K^{2}(x,y)\dd x\dd y+2\int_{0}^{\infty}\!\!\int_{0}^{\infty}\!\!\int_{0}^{\infty}\!\!xyz\times \nonumber \\
\fl&& \ln x\ln y\ln z~K(x,y)K(y,z)K(z,x)\dd x\dd y\dd z.
\end{eqnarray}

We now turn to the third cumulant of the induced entropy $T$, which will result in a more compact expression~(\ref{eq:T3ABC}) than that of the corresponding moment~(\ref{eq:T3}). Similarly as the third moment~(\ref{eq:T3}), the first two moments of $T$ can also be represented as integrals involving the correlation kernel as
\begin{eqnarray}
\mathbb{E}_{g}\!\left[T\right]&=&\int_{0}^{\infty}\!\!x\ln{x}~K(x,x)\dd{x} \label{eq:T1} \\
\mathbb{E}_{g}\!\left[T^{2}\right]&=&\left(\int_{0}^{\infty}\!\!x\ln{x}~K(x,x)\dd{x}\right)^{2}+\int_{0}^{\infty}\!\!x^{2}\ln^{2}x~K(x,x)\dd{x}-\nonumber\\
&&\int_{0}^{\infty}\!\!\int_{0}^{\infty}\!\!xy\ln x\ln y~K^{2}(x,y)\dd x\dd y. \label{eq:T2}
\end{eqnarray}
The above integrals have been computed in~\cite{Ruiz95} and~\cite{Wei17}, respectively, as
\begin{eqnarray}
\mathbb{E}_{g}\!\left[T\right]&=&mn\psi_{0}(n)+\frac{1}{2}m(m+1) \label{eq:T1R} \\
\mathbb{E}_{g}\!\left[T^{2}\right]&=&mn(m+n)\psi_{1}(n)+mn(mn+1)\psi_{0}^{2}(n)+m\big(m^{2}n+mn+ \nonumber\\
&&m+2n+1\big)\psi_{0}(n)+\frac{1}{4}m(m+1)\left(m^2+m+2\right). \label{eq:T2R}
\end{eqnarray}
Inserting the first three moments~(\ref{eq:T1}),~(\ref{eq:T2}), and~(\ref{eq:T3}) into the moment-to-cumulant relation~(\ref{eq:3mk}), the third cumulant of $T$ is simplified to
\begin{eqnarray}
\kappa_{3}^{T}&=&\mathbb{E}_{g}\!\left[T^{3}\right]-3\mathbb{E}_{g}\!\left[T^{2}\right]\mathbb{E}_{g}\!\left[T\right]+2\mathbb{E}_{g}^{3}\!\left[T\right]\\
&=&I_{A}-3I_{B}+2I_{C}, \label{eq:T3ABC}
\end{eqnarray}
where we denote the integrals
\begin{eqnarray}
I_{A} &=& \int_{0}^{\infty}\!\!x^{3}\ln^{3}{x}~K(x,x)\dd{x} \label{eq:IA} \\
I_{B} &=& \int_{0}^{\infty}\!\!\int_{0}^{\infty}\!\!x^{2}y\ln^{2}x\ln y~K^{2}(x,y)\dd x\dd y \label{eq:IB} \\
I_{C} &=& \int_{0}^{\infty}\!\!\int_{0}^{\infty}\!\!\int_{0}^{\infty}\!\!xyz\ln x\ln y\ln z~K(x,y)K(y,z)K(z,x)\dd x\dd y\dd z. \label{eq:IC}
\end{eqnarray}
By the relations~(\ref{eq:1km})-(\ref{eq:3mk}) and the existing results~(\ref{eq:T1R}),~(\ref{eq:T2R}), it remains to compute the integrals $I_{A}$, $I_{B}$, and $I_{C}$ in order to obtain the final result~(\ref{eq:k3}). The prior results discussed so far also allow us to express the desired third cumulant of $S$ in terms of the first three cumulants of $T$ as
\\
\begin{eqnarray}\label{eq:k3T3}
\kappa_{3}&=&\frac{1}{(mn)_{3}}\Bigg(\!-\kappa_{3}^{T}+\frac{6}{mn}\kappa_{1}^{T}\kappa_{2}^{T}+\frac{3(2mn+3)}{mn+1}\kappa_{2}^{T}-\frac{4}{m^{2}n^{2}}\left(\kappa_{1}^{T}\right)^{3}-\nonumber\\
&&\frac{3(3mn+4)}{mn(mn+1)}\left(\kappa_{1}^{T}\right)^{2}-\frac{6(mn+2)}{mn+1}\kappa_{1}^{T}\Bigg)+\psi_{2}(mn+1),
\end{eqnarray}
where~(\ref{eq:T1R}) and~(\ref{eq:T2R}) directly lead to
\begin{eqnarray}
\fl\kappa_{1}^{T}&=&mn\psi_{0}(n)+\frac{1}{2}m(m+1) \label{eq:kT1} \\
\fl\kappa_{2}^{T}&=&mn(m+n)\psi_{1}(n)+mn\psi_{0}^{2}(n)+m\left(m+2n+1\right)\psi_{0}(n)+\frac{1}{2}m(m+1), \label{eq:kT2}
\end{eqnarray}
and we have also utilized the identities below, cf.~(\ref{eq:pg}), in simplifying~(\ref{eq:k3T3}),
\numparts
\begin{eqnarray}
\psi_{0}(l+n)&=&\psi_{0}(l)+\sum_{k=0}^{n-1}\frac{1}{l+k} \label{eq:0s} \\
\psi_{1}(l+n)&=&\psi_{1}(l)-\sum_{k=0}^{n-1}\frac{1}{(l+k)^2} \label{eq:1s} \\
\psi_{2}(l+n)&=&\psi_{2}(l)+2\sum_{k=0}^{n-1}\frac{1}{(l+k)^3}. \label{eq:2s}
\end{eqnarray}
\endnumparts
The rest of the paper is to compute the integrals $I_{A}$, $I_{B}$, $I_{C}$, where we will eventually show that
\begin{eqnarray}
\kappa_{3}^{T}&=&I_{A}-3I_{B}+2I_{C} \label{eq:ABC} \\
&=&mn\left(m^2+3mn+n^2+1\right)\psi_{2}(n)+6mn(m+n)\psi_{0}(n)\psi_{1}(n)+\nonumber\\
&&m\left(2m^2+12mn+3m+6n^{2}+3n+1\right)\psi_{1}(n)+2mn\psi_{0}^{3}(n)+\nonumber\\
&&3m(m+3n+1)\psi_{0}^{2}(n)+6m(m+n+1)\psi_{0}(n)+m(m+1). \label{eq:T3R}
\end{eqnarray}
Inserting the above expression as well as~(\ref{eq:kT1}) and~(\ref{eq:kT2}) into~(\ref{eq:k3T3}), the claimed main result~(\ref{eq:k3}) will then be established.

\subsection{Calculations of integrals $I_{A}$, $I_{B}$, and $I_{C}$}\label{sec:IABC}
To compute the remaining integrals~(\ref{eq:IA}),~(\ref{eq:IB}), and~(\ref{eq:IC}), we need the following results on the Wishart-Laguerre ensemble. First, its correlation kernel can be explicitly written as~\cite{Forrester}
\begin{equation}\label{eq:ker}
K(x_{i},x_{j})=\sqrt{\e^{-x_{i}-x_{j}}(x_{i}x_{j})^{n-m}}\sum_{k=0}^{m-1}\frac{C_{k}(x_{i})C_{k}(x_{j})}{k!(n-m+k)!},
\end{equation}
where
\begin{equation}
C_{k}(x)=(-1)^{k}k!L_{k}^{(n-m)}(x)
\end{equation}
with
\begin{equation}\label{eq:Lar}
L_{k}^{(n-m)}(x)=\sum_{i=0}^{k}(-1)^{i}{n-m+k\choose k-i}\frac{x^i}{i!}
\end{equation}
being the (generalized) Laguerre polynomial of degree $k$. Similarly as the Hermite polynomials~(\ref{eq:ocH}), the Laguerre polynomials also satisfy an orthogonality relation~\cite{Forrester}
\begin{equation}\label{eq:oc}
\int_{0}^{\infty}\!\!x^{n-m}\e^{-x}L_{k}^{(n-m)}(x)L_{l}^{(n-m)}(x)\dd{x}=\frac{(n-m+k)!}{k!}\delta_{kl}.
\end{equation}
Instead of the summation form~(\ref{eq:ker}), the one arbitrary eigenvalue density~(\ref{eq:g1}) admits a more convenient form by the Christoffel-Darboux formula~\cite{Ruiz95,Forrester} as
\begin{equation}\label{eq:one}
\!\!\!\!\!\!\!\!\!\!\!\!\!\!\!\!\!\!\!\!g_{1}(x)=\frac{(m-1)!}{(n-1)!}x^{n-m}\e^{-x}\left(\left(L_{m-1}^{(n-m+1)}(x)\right)^{2}-L_{m-2}^{(n-m+1)}(x)L_{m}^{(n-m+1)}(x)\right).
\end{equation}
We also need the following integral, due to Schr{\"{o}}dinger~\cite{Schrodinger1926}, that generalizes the identity~(\ref{eq:oc}) to
\begin{equation}\label{eq:Swm}
\fl\int_{0}^{\infty}\!\!x^{q}\e^{-x}L_{s}^{(\alpha)}(x)L_{t}^{(\beta)}(x)\dd{x}=(-1)^{s+t}\sum_{k=0}^{\min(s,t)}{q-\alpha\choose s-k}{q-\beta\choose t-k}\frac{\Gamma(q+1+k)}{k!}.
\end{equation}
By taking up to the third derivatives of the Schr{\"{o}}dinger's integral~(\ref{eq:Swm}) with respect to $q$, we obtain three more useful integrals shown in~(\ref{eq:1d}) (see also~\cite{Ruiz95}),~(\ref{eq:2d}), and~(\ref{eq:3d}) below, denoted respectively by $A_{s,t}^{(\alpha,\beta)}(q)$, $B_{s,t}^{(\alpha,\beta)}(q)$, and $C_{s,t}^{(\alpha,\beta)}(q)$, where we have also denoted
\begin{eqnarray}
\Psi_{j}&=&\psi_{j}(q+1+k)+\psi_{j}(q-\alpha+1)+\psi_{j}(q-\beta+1)- \nonumber \\
&&\psi_{j}(q-\alpha-s+1+k)-\psi_{j}(q-\beta-t+1+k).
\end{eqnarray}
Note that the explicit kernel expression~(\ref{eq:ker}) together with the corresponding Schr{\"{o}}dinger's integral~(\ref{eq:Swm}) (and its higher order derivatives) of the Wishart-Laguerre ensemble, unavailable for the fixed-trace ensemble, makes the subsequent calculation possible. This fact is the motivation behind the approach of moments (cumulants) conversion~(\ref{eq:mST}) between the two ensembles.

\begin{eqnarray}
\fl A_{s,t}^{(\alpha,\beta)}(q)&=&\int_{0}^{\infty}\!\!x^{q}\e^{-x}\ln{x}~L_{s}^{(\alpha)}(x)L_{t}^{(\beta)}(x)\dd{x} \label{eq:i1d}\\
\fl&=&(-1)^{s+t}\sum_{k=0}^{\min(s,t)}{q-\alpha\choose s-k}{q-\beta\choose t-k}\frac{\Gamma(q+1+k)}{k!}\Psi_{0}.\label{eq:1d}
\end{eqnarray}

\begin{eqnarray}
\fl B_{s,t}^{(\alpha,\beta)}(q)&=&\int_{0}^{\infty}\!\!x^{q}\e^{-x}\ln^{2}{x}~L_{s}^{(\alpha)}(x)L_{t}^{(\beta)}(x)\dd{x} \label{eq:i2d}\\
\fl&=&(-1)^{s+t}\sum_{k=0}^{\min(s,t)}{q-\alpha\choose s-k}{q-\beta\choose t-k}\frac{\Gamma(q+1+k)}{k!}\left(\Psi_{0}^{2}+\Psi_{1}\right).\label{eq:2d}
\end{eqnarray}

\begin{eqnarray}
\fl C_{s,t}^{(\alpha,\beta)}(q)&=&\int_{0}^{\infty}\!\!x^{q}\e^{-x}\ln^{3}{x}~L_{s}^{(\alpha)}(x)L_{t}^{(\beta)}(x)\dd{x}\label{eq:i3d}\\
\fl&=&(-1)^{s+t}\sum_{k=0}^{\min(s,t)}{q-\alpha\choose s-k}{q-\beta\choose t-k}\frac{\Gamma(q+1+k)}{k!}\left(\Psi_{0}^{3}+3\Psi_{0}\Psi_{1}+\Psi_{2}\right).\label{eq:3d}
\end{eqnarray}

\subsubsection{Computing $I_{A}$}
Inserting the one arbitrary eigenvalue density~(\ref{eq:one}) (see also~(\ref{eq:g1})) into~(\ref{eq:IA}), $I_{A}$ is expressed via the integral~(\ref{eq:i3d}) as
\begin{equation}\label{eq:IAI}
\!\!\!\!\!\!\!\!\!\!\!\!\!\!\!\!\!\!\!\!\!\!\!\!\!\!I_{A}=\frac{m!}{(n-1)!}\left(C_{m-1,m-1}^{(n-m+1,n-m+1)}(n-m+3)-C_{m-2,m}^{(n-m+1,n-m+1)}(n-m+3)\right),
\end{equation}
where the integrals in~(\ref{eq:IAI}) are evaluated by invoking the identity~(\ref{eq:3d}) and then collecting the non-zero contributions. In particular, the non-zero contributions consist of indeterminate terms as a result of zero or negative arguments of gamma and polygamma functions. These indeterminacy can be resolved by interpreting the gamma and polygamma functions involved as the limits $\epsilon\to0$ of (with $l\geq0$)
\numparts
\label{eq:pgna}\begin{eqnarray}
\!\!\!\!\!\!\!\!\!\!\!\!\!\!\!\!\!\!\!\!\!\!\!\!\!\!\!\!\!\Gamma(-l+\epsilon)&=&\frac{(-1)^{l}}{l!\epsilon}\left(1+\psi_{0}(l+1)\epsilon+o\left(\epsilon^2\right)\right)\label{eq:pgna1}\\
\!\!\!\!\!\!\!\!\!\!\!\!\!\!\!\!\!\!\!\!\!\!\!\!\!\!\!\!\!\psi_{0}(-l+\epsilon)&=&-\frac{1}{\epsilon}+\psi_{0}(l+1)+\left(2\psi_{1}(1)-\psi_{1}(l+1)\right)\epsilon+\frac{1}{2}\psi_{2}(l+1)\epsilon^{2}+o\left(\epsilon^3\right)\label{eq:pgna2}\\
\!\!\!\!\!\!\!\!\!\!\!\!\!\!\!\!\!\!\!\!\!\!\!\!\!\!\!\!\!\psi_{1}(-l+\epsilon)&=&\frac{1}{\epsilon^{2}}-\psi_{1}(l+1)+\psi_{1}(1)+\zeta(2)+o\left(\epsilon\right)\label{eq:pgna3}\\
\!\!\!\!\!\!\!\!\!\!\!\!\!\!\!\!\!\!\!\!\!\!\!\!\!\!\!\!\!\psi_{2}(-l+\epsilon)&=&-\frac{2}{\epsilon^{3}}+\psi_{2}(l+1)+\psi_{2}(1)+2\zeta(3)+o\left(\epsilon\right).\label{eq:pgna4}
\end{eqnarray}
\endnumparts
By the above procedure, the integrals $C_{m-1,m-1}^{(n-m+1,n-m+1)}(n-m+3)$ and $C_{m-2,m}^{(n-m+1,n-m+1)}(n-m+3)$ are evaluated, respectively, as
\begin{eqnarray}
\fl&&C_{m-1,m-1}^{(n-m+1,n-m+1)}(n-m+3)\\
\fl=&&\int_{0}^{\infty}\!\!x^{n-m+3}\e^{-x}\ln^{3}{x}~\left(L_{m-1}^{(n-m+1)}(x)\right)^{2}\dd{x}\\
\fl=&&\frac{(n-1)!}{2(m-1)!}\Big(18m^{2}n+39m^2-30mn-57m+12n+30+3\big(13m^{2}n+12m^2+4mn^2-\nonumber\\
\fl&&3mn-4m-4n^2+2n+4\big)\psi_{0}(n)+6\left(3m^{2}n+m^2+4mn^2+3mn+m-n^2\right)\times\nonumber\\
\fl&&\left(\psi_{0}^{2}(n)+\psi_{1}(n)\right)+2n\left(m^2+4mn+m+n^2-n\right)\!\left(\psi_{0}^{3}(n)+3\psi_{0}(n)\psi_{1}(n)+\psi_{2}(n)\right)\!\Big)+\nonumber\\
\fl&&\sum_{k=1}^{m-3}\frac{6(n-k)!}{(m-3-k)!}\left(\frac{3}{k+2}-\frac{3}{k}+\frac{1}{k^2}+\frac{4}{(k+1)^2}+\frac{1}{(k+2)^2}\right)\big(\psi_{0}(n+1-k)-\nonumber\\
\fl&&2\psi_{0}(k)+2\psi_{0}(1)+3\big),\label{eq:IAS1}
\end{eqnarray}
and
\begin{eqnarray}
\fl&&C_{m-2,m}^{(n-m+1,n-m+1)}(n-m+3)\\
\fl=&&\int_{0}^{\infty}\!\!x^{n-m+3}\e^{-x}\ln^{3}{x}~L_{m-2}^{(n-m+1)}(x)L_{m}^{(n-m+1)}(x)\dd{x}\\
\fl=&&\frac{(n-1)!}{8(m-2)!}\Big(40mn-m^2+69m-32n-38+8\big(8mn-m^2+9m+3n^2+5n+1)\times\nonumber\\
\fl&&\psi_{0}(n)+2\left(8mn-m^2+5m+18n^2+26n+6\right)\left(\psi_{0}^{2}(n)+\psi_{1}(n)\right)+8n(n+1)\times\nonumber\\
\fl&&\left(\psi_{0}^{3}(n)+3\psi_{0}(n)\psi_{1}(n)+\psi_{2}(n)\right)+\sum_{k=1}^{m-4}\frac{(n-k-1)!}{(m-k-4)!}\Bigg(\frac{1}{2k}-\frac{4}{k+1}+\frac{4}{k+3}-\nonumber\\
\fl&&\frac{1}{2(k+4)}+\frac{6}{(k+2)^2}\Bigg)\left(\psi_{0}(n-k)-\psi_{0}(k+2)-\psi_{0}(k)+2\psi_{0}(1)+3\right).\label{eq:IAS2}
\end{eqnarray}
Inserting~(\ref{eq:IAS1}) and~(\ref{eq:IAS2}) back into~(\ref{eq:IAI}), we arrive at a finite summation form of $I_{A}$.


\subsubsection{Computing $I_{B}$}
Similarly, inserting the kernel expression~(\ref{eq:ker}) into~(\ref{eq:IB}), $I_{B}$ is expressed via the integrals~(\ref{eq:i1d}) and~(\ref{eq:i2d}) as
\begin{eqnarray}\label{eq:IBI}
\fl I_{B}&=&\sum_{k=0}^{m-1}\frac{k!^{2}A_{k,k}^{(n-m,n-m)}(n-m+1)B_{k,k}^{(n-m,n-m)}(n-m+2)}{(k+n-m)!^{2}}+\nonumber\\
\fl&&\sum_{k=0}^{m-2}\sum_{j=0}^{m-k-2}\frac{2j!(k+j+1)!A_{j,k+j+1}^{(n-m,n-m)}(n-m+1)B_{j,k+j+1}^{(n-m,n-m)}(n-m+2)}{(j+n-m)!(k+j+n-m+1)!}.
\end{eqnarray}
By the identities~(\ref{eq:1d}),~(\ref{eq:2d}) and with the help of~$(106)$, the integrals $A_{k,k}^{(n-m,n-m)}(n-m+1)$ and $B_{k,k}^{(n-m,n-m)}(n-m+2)$ in~(\ref{eq:IBI}) are calculated as
\begin{eqnarray}
\fl&&A_{k,k}^{(n-m,n-m)}(n-m+1)\\
\fl=&&\int_{0}^{\infty}\!\!x^{n-m+1}\e^{-x}\ln{x}~\left(L_{k}^{(n-m)}(x)\right)^{2}\dd{x} \\
\fl=&&\frac{(k+n-m)!}{k!}\left((2k-m+n+1)\psi_{0}(k+n-m+1)+2k+1\right),\label{eq:IBS1}
\end{eqnarray}

\begin{eqnarray}
\fl&&B_{k,k}^{(n-m,n-m)}(n-m+2)\\
\fl=&&\int_{0}^{\infty}\!\!x^{n-m+2}\e^{-x}\ln^{2}{x}~\left(L_{k}^{(n-m)}(x)\right)^{2}\dd{x} \\
\fl=&&\frac{(k+n-m)!}{k!}\Bigg(\frac{1}{2}\left(17k^2+k(4n-4m+7)+4\right)+2\big(7k^2+k(4n-4m+7)+2n- \nonumber\\
\fl&&2m+3\big)\psi_{0}(k+n-m+1)+\left(6k^2+6k(n-m+1)+(n-m+2)(n-m+1)\right)\times \nonumber\\
\fl&&\left(\psi_{0}^{2}(k+n-m+1)+\psi_{1}(k+n-m+1)\right)+\sum_{i=3}^{k}\frac{2(k-i+n-m+2)!}{(k-i)!}\times \nonumber\\
\fl&&\left(\frac{3}{i}-\frac{3}{i-2}+\frac{1}{i^2}+\frac{4}{(i-1)^2}+\frac{1}{(i-2)^2}\right)\!\Bigg),\label{eq:IBS2}
\end{eqnarray}
and the integrals $A_{j,k+j+1}^{(n-m,n-m)}(n-m+1)$ and $B_{j,k+j+1}^{(n-m,n-m)}(n-m+2)$ admit different expressions for different ranges of $k$ as
\begin{eqnarray}
\fl&&A_{j,j+1}^{(n-m,n-m)}(n-m+1)\\
\fl&=&\int_{0}^{\infty}\!\!x^{n-m+1}\e^{-x}\ln{x}~L_{j}^{(n-m)}(x)L_{j+1}^{(n-m)}(x)\dd{x} \\
\fl&=&-\frac{(j+n-m)!}{j!}\left((j+n-m+1)\psi_{0}(j+n-m+1)+\frac{3j}{2}+n-m+2\right),\label{eq:IBS3}
\end{eqnarray}

\begin{eqnarray}
\fl&&A_{j,k+j+1}^{(n-m,n-m)}(n-m+1)\\
\fl=&&\int_{0}^{\infty}\!\!x^{n-m+1}\e^{-x}\ln{x}~L_{j}^{(n-m)}(x)L_{k+j+1}^{(n-m)}(x)\dd{x} \\
\fl=&&\frac{(j+n-m)!}{j!(k+1)}\left(\frac{j+n-m+1}{k}-\frac{j}{k+2}\right),~~~~~~k>0,\label{eq:IBS4}
\end{eqnarray}

\begin{eqnarray}
\fl&&B_{j,j+1}^{(n-m,n-m)}(n-m+2)\\
\fl=&&\int_{0}^{\infty}\!\!x^{n-m+2}\e^{-x}\ln^{2}{x}~L_{j}^{(n-m)}(x)L_{j+1}^{(n-m)}(x)\dd{x} \\
\fl=&&-\frac{2(j+n-m)!}{j!}\Bigg(\frac{7j^{2}}{2}+\frac{5j}{2}(n-m+3)+2n-2m+5+\Bigg(\frac{16j^{2}}{3}+j\Bigg(6n-6m+\nonumber\\
\fl&&\frac{38}{3}\Bigg)+(n-m)^2+7(n-m)+8\Bigg)\psi_{0}(j+n-m+1)+(j+n-m+1)(2j+n-\nonumber\\
\fl&&m+2)\left(\psi_{0}^{2}(j+n-m+1)+\psi_{1}(j+n-m+1)\right)\!\Bigg)+\sum_{l=3}^{j}\frac{2(j-l+n-m+2)!}{(j-l)!}\times\nonumber\\
\fl&&\left(-\frac{1}{3(l+1)}-\frac{3}{l}+\frac{3}{l-1}+\frac{1}{3(l-2)}-\frac{2}{l^2}-\frac{2}{(l-1)^2}\right),\label{eq:IBS5}
\end{eqnarray}

\begin{eqnarray}
\fl&&B_{j,j+2}^{(n-m,n-m)}(n-m+2)\\
\fl=&&\int_{0}^{\infty}\!\!x^{n-m+2}\e^{-x}\ln^{2}{x}~L_{j}^{(n-m)}(x)L_{j+2}^{(n-m)}(x)\dd{x} \\
\fl=&&\frac{(j+n-m)!}{j!}\Bigg(\frac{10j^2}{3}+\frac{2j}{3}(7n-7m+20)+(n-m)^2+9(n-m)+13+\frac{1}{6}\big(25j^2+\nonumber\\
\fl&&j(44n-44m+87)+6(n-m+3)(3n-3m+4)\big)\psi_{0}(j+n-m+1)+(j+n-\nonumber\\
\fl&&m+1)(j+n-m+2)\left(\psi_{0}^{2}(j+n-m+1)+\psi_{1}(j+n-m+1)\right)\!\Bigg)+\nonumber\\
\fl&&\sum_{l=3}^{j}\!\frac{4(j-l+n-m+2)!}{(j-l)!}\left(\frac{1}{3(l+1)}-\frac{1}{24(l+2)}-\frac{1}{3(l-1)}+\frac{1}{24(l-2)}+\frac{1}{2l^2}\right)\!,\label{eq:IBS6}
\end{eqnarray}

\begin{eqnarray}
\fl&&B_{j,j+k+1}^{(n-m,n-m)}(n-m+2)\\
\fl=&&\int_{0}^{\infty}\!\!x^{n-m+2}\e^{-x}\ln^{2}{x}~L_{j}^{(n-m)}(x)L_{j+k+1}^{(n-m)}(x)\dd{x} \\
\fl=&&-\frac{2(j+n-m)!}{j!(k-1)_{5}}\Bigg(j^{2}\left(k^2-7k+24\psi_{0}(1)+56\right)+j\big(k^{2}(2m-2n-5)+k(18n-\nonumber\\
\fl&&18m+55+12(n-m+2)\psi_{0}(1))+4(3(3n-3m+4)\psi_{0}(1)+18n-18n+31)\big)+\nonumber\\
\fl&&(k+2)(k+3)\big(3m^2-6mn-13m+3n^2+13n+12+2(n-m+1)(n-m+2)\times\nonumber\\
\fl&&\psi_{0}(1)\big)+2\big(12j^2+6j(k(n-m+2)+3n-3m+4)+(k+2)(k+3)(n-m+1)\times\nonumber\\
\fl&&(n-m+2)\big)\left(\psi_{0}(j+n-m+1)-\psi_{0}(k-1)\right)\!\Bigg)+\sum_{l=3}^{j}\frac{8(j-l+n-m+2)!}{(j-l)!}\times\nonumber\\
\fl&&\frac{1}{(l-2)(l-1)l(k+l-1)(k+l)(k+l+1)},~~~~~~k>1.\label{eq:IBS7}
\end{eqnarray}
Inserting~(\ref{eq:IBS1}),~(\ref{eq:IBS2}),~(\ref{eq:IBS3}),~(\ref{eq:IBS4}),~(\ref{eq:IBS5}),~(\ref{eq:IBS6}), and~(\ref{eq:IBS7}) back into~(\ref{eq:IBI}), we obtain a finite summation representation of $I_{B}$.

\subsubsection{Computing $I_{C}$}
In the same manner, by inserting~(\ref{eq:ker}) into~(\ref{eq:IC}), $I_{C}$ can be expressed in terms of the integral~(\ref{eq:i1d}) as
\begin{eqnarray}\label{eq:ICI}
\fl
I_{C}&=&\sum_{k=0}^{m-1}\frac{k!^{3}\left(A_{k,k}^{(n-m,n-m)}(n-m+1)\right)^{3}}{(k+n-m)!^{3}}+\sum_{j=0}^{m-1}\sum_{i=j+1}^{m-1}\frac{3i!j!\left(A_{i,j}^{(n-m,n-m)}(n-m+1)\right)^{2}}{(i+n-m)!(j+n-m)!}\times\nonumber\\
\fl&&\left(\frac{i!A_{i,i}^{(n-m,n-m)}(n-m+1)}{(i+n-m)!}+\frac{j!A_{j,j}^{(n-m,n-m)}(n-m+1)}{(j+n-m)!}\right)+\sum_{k=0}^{m-1}\sum_{j=k+1}^{m-1}\sum_{i=j+1}^{m-1}6\times\nonumber\\
\fl&&\frac{i!j!k!A_{i,j}^{(n-m,n-m)}(n-m+1)A_{j,k}^{(n-m,n-m)}(n-m+1)A_{k,i}^{(n-m,n-m)}(n-m+1)}{(i+n-m)!(j+n-m)!(k+n-m)!}.
\end{eqnarray}
The integrals in the above expression have been computed in~(\ref{eq:IBS1}),~(\ref{eq:IBS3}), and~(\ref{eq:IBS4}), inserting which back into~(\ref{eq:ICI}) leads to a finite summation form of $I_{C}$.

\subsection{Evaluation of summations in $I_{A}$, $I_{B}$, and $I_{C}$}\label{sec:simp}
The rest of the task is to evaluate the summations in the finite summation forms of $I_{A}$, $I_{B}$, and $I_{C}$, after inserting back the computed  integrals~(\ref{eq:IAS1}), (\ref{eq:IAS2}), (\ref{eq:IBS1}), (\ref{eq:IBS2}), (\ref{eq:IBS3}), (\ref{eq:IBS4}), (\ref{eq:IBS5}), (\ref{eq:IBS6}), (\ref{eq:IBS7}). The evaluation of the summations amounts to applying the closed-form identities listed in~\ref{sec:ap1-c} and~\ref{sec:ap2-c} as well as the semi closed-form identities listed in~\ref{sec:ap1-sc} and~\ref{sec:ap2-sc}. The derivation of these identities is discussed in~\ref{sec:ap1-re} and~\ref{sec:ap2-re}.

Evaluating the summations is a tedious but straightforward procedure, which eventually leads to $I_{A}$, $I_{B}$, and $I_{C}$ being expressed respectively by~(\ref{eq:IAf}),~(\ref{eq:IBf}), and~(\ref{eq:ICf}) as shown below. The coefficients $a_{i}$, $b_{i}$, and $c_{i}$ in the expressions~(\ref{eq:IAf}),~(\ref{eq:IBf}), and~(\ref{eq:ICf}) can be found in Table~\ref{t:IAc}, Table~\ref{t:IBc}, and Table~\ref{t:ICc}, respectively. As seen from~(\ref{eq:IAf}),~(\ref{eq:IBf}), and~(\ref{eq:ICf}), we have shifted the argument of each polygamma function in $I_{A}$, $I_{B}$, and $I_{C}$ to one of following: $1$, $m$, $n$, $n-m$, $n-m+k$, with the help of~(89). This choice leads the coefficients in Table~\ref{t:IAc}, Table~\ref{t:IBc}, and Table~\ref{t:ICc}, to become polynomials in $m$ and $n$.

\begin{eqnarray}
\fl I_{A}&=&a_{1}+a_{2}\psi_{0}(n)+a_{3}\psi_{0}(n-m)+a_{4}\psi_{0}(1)\psi_{0}(n-m)+a_{5}\psi_{0}(m)\psi_{0}(n-m)+\nonumber\\
\fl&&a_{6}\psi_{0}(n)\psi_{0}(n-m)+a_{7}\psi_{0}^{2}(n-m)+a_{8}\psi_{0}(1)\psi_{0}(n)\psi_{0}(n-m)+\nonumber\\
\fl&&a_{9}\psi_{0}(1)\psi_{0}^{2}(n-m)+a_{10}\psi_{0}(m)\psi_{0}(n)\psi_{0}(n-m)+a_{11}\psi_{0}(m)\psi_{0}^{2}(n-m)+\nonumber\\
\fl&&a_{12}\psi_{0}(n)\psi_{0}^{2}(n-m)+a_{13}\psi_{0}^{3}(n-m)+a_{14}\psi_{1}(n-m)+a_{15}\psi_{0}(n)\psi_{1}(n-m)+\nonumber\\
\fl&&a_{16}\psi_{2}(n-m)+a_{17}\sum_{k=1}^m\frac{\psi_{0}(k+n-m)}{k}+a_{18}\sum_{k=1}^{m}\frac{\psi_{0}^{2}(k+n-m)}{k}.\label{eq:IAf}
\end{eqnarray}

\begin{longtable}[h!]{cl}
\caption{Coefficients in~(\ref{eq:IAf}) of $I_{A}$.}\\
\ns\ns\ns
\hline\hline

\bs\multirow{2}{*}{$a_{1}=$} & $\frac{m}{288}\big(37 m^3+4012 m^2 n-30 m^2+4410 m n^2-330 m n-169 m+84 n^3-$ \\ \bs
& $30 n^2-250 n+162\big)$ \\ \bs

\bs\multirow{2}{*}{$a_{2}=$} & $-\frac{n}{24}\big(12 m^3+414 m^2 n-6 m^2+364 m n^2+6 m n-94 m+7 n^3-6 n^2-$ \\ \bs
& $67 n+90\big)$ \\ \bs

\bs\multirow{2}{*}{$a_{3}=$} & $-\frac{1}{24}\big(7 m^4+352 m^3 n+18 m^3+336 m^2 n+5 m^2-352 m n^3+360 m n^2+$ \\ \bs
& $216 m n+42 m-7 n^4+6 n^3+139 n^2+222 n+72\big)$ \\ \bs

\bs$a_{4}=$ & $\frac{m}{2}\left(m^3+28 m^2 n+6 m^2+30 m n^2+18 m n+11 m+6 n^2+26 n+6\right)$ \\ \bs

\bs$a_{5}=$ & $-\frac{m}{2}\left(m^3+28 m^2 n+6 m^2+30 m n^2+18 m n+11 m+6 n^2+26 n+6\right)$ \\ \bs

\bs$a_{6}=$ & $\frac{n}{2}\left(30 m^2 n-18 m^2+28 m n^2-54 m n+26 m+n^3-18 n^2+11 n-18\right)$ \\ \bs

\bs\multirow{2}{*}{$a_{7}=$} & $\frac{1}{4}\big(m^4+28 m^3 n+6 m^3-30 m^2 n^2+6 m^2 n+11 m^2-56 m n^3-$ \\ \bs
& $30 m n^2-26 m n+6 m-2 n^4-12 n^3-22 n^2-12 n\big)$ \\ \bs

\bs$a_{8}=$ & $6 m n \left(m^2+3 m n+n^2+1\right)$ \\ \bs

\bs$a_{9}=$ & $6 m n \left(m^2+3 m n+n^2+1\right)$ \\ \bs

\bs$a_{10}=$ & $-6 m n \left(m^2+3 m n+n^2+1\right)$ \\ \bs

\bs$a_{11}=$ & $-6 m n \left(m^2+3 m n+n^2+1\right)$ \\ \bs

\bs$a_{12}=$ & $3 m n \left(m^2+3 m n+n^2+1\right)$ \\ \bs

\bs$a_{13}=$ & $-2 m n \left(m^2+3 m n+n^2+1\right)$ \\ \bs

\bs$a_{14}=$ & $\frac{m}{4}\left(m^3+28 m^2 n+6 m^2+30 m n^2+18 m n+11 m+6 n^2+26 n+6\right)$ \\ \bs

\bs$a_{15}=$ & $3 m n \left(m^2+3 m n+n^2+1\right)$ \\ \bs

\bs$a_{16}=$ & $m n \left(m^2+3 m n+n^2+1\right)$ \\ \bs

\bs\multirow{2}{*}{$a_{17}=$} & $\frac{m}{2}\big(m^3+28 m^2 n+6 m^2+30 m n^2+18 m n+11 m+6 n^2+26 n+6+$ \\ \bs
& $12n\left(m^2+3 m n+n^2+1\right)\psi_{0}(n)+24n\left(m^2+3 m n+n^2+1\right)\psi_{0}(n-m)\big)$ \\ \bs

\bs$a_{18}=$ & $-6 m n \left(m^2+3 m n+n^2+1\right)$ \\ \bs

\hline\hline\bs
\caption{Coefficients in~(\ref{eq:IAf}) of $I_{A}$.} \\
\label{t:IAc}
\end{longtable}

\begin{eqnarray}
\fl I_{B}&=&b_{1}+b_{2}\psi_{0}(n)+b_{3}\psi_{0}(n-m)+b_{4}\psi_{0}(1)\psi_{0}(n-m)+b_{5}\psi_{0}(m)\psi_{0}(n-m)+\nonumber\\
\fl&&b_{6}\psi_{0}^{2}(n)+b_{7}\psi_{0}(n)\psi_{0}(n-m)+b_{8}\psi_{0}^{2}(n-m)+b_{9}\psi_{0}(1)\psi_{0}(n)\psi_{0}(n-m)+\nonumber\\
\fl&&b_{10}\psi_{0}(1)\psi_{0}^{2}(n-m)+b_{11}\psi_{0}(m)\psi_{0}(n)\psi_{0}(n-m)+b_{12}\psi_{0}^{2}(n)\psi_{0}(n-m)+\nonumber\\
\fl&&b_{13}\psi_{0}(m)\psi_{0}^{2}(n-m)+b_{14}\psi_{0}(n)\psi_{0}^{2}(n-m)+b_{15}\psi_{0}^{3}(n-m)+b_{16}\psi_{1}(n)+\nonumber\\
\fl&&b_{17}\psi_{0}(n-m)\psi_{1}(n)+b_{18}\psi_{1}(n-m)+b_{19}\psi_{0}(1)\psi_{1}(n-m)+\nonumber\\
\fl&&b_{20}\psi_{0}(m)\psi_{1}(n-m)+b_{21}\psi_{0}(n)\psi_{1}(n-m)+b_{22}\psi_{0}(n-m)\psi_{1}(n-m)+\nonumber\\
\fl&&b_{23}\sum_{k=1}^{m}\frac{\psi_{0}(k+n-m)}{k}+b_{24}\sum_{k=1}^{m}\frac{\psi_{0}^{2}(k+n-m)}{k}+b_{25}\sum_{k=1}^{m}\frac{\psi_{1}(k+n-m)}{k}.\label{eq:IBf}
\end{eqnarray}

\begin{longtable}[h!]{cl}
\caption{Coefficients in~(\ref{eq:IBf}) of $I_{B}$.}\\
\ns\ns\ns
\hline\hline

\bs\multirow{2}{*}{$b_{1}=$} & $\frac{m}{864}\big(111 m^3+12036 m^2 n+2198 m^2+13230 m n^2+4530 m n+1629 m+$ \\ \bs
& $252 n^3+150 n^2-414 n-194\big)$ \\ \bs

\bs\multirow{2}{*}{$b_{2}=$} & $-\frac{1}{72}\big(36 m^3 n+1242 m^2 n^2+174 m^2 n+72 m^2+1092 m n^3+690 m n^2-$ \\ \bs
& $234 m n+168 m+21 n^4+14 n^3-321 n^2+358 n+24\big)$ \\ \bs

\bs\multirow{2}{*}{$b_{3}=$} & $-\frac{1}{72}\big(21 m^4+1056 m^3 n+230 m^3+864 m^2 n+255 m^2-1056 m n^3+$ \\ \bs
& $360 m n^2+216 m n+94 m-21 n^4-14 n^3+249 n^2+434 n+144\big)$ \\ \bs

\bs$b_{4}=$ & $\frac{m}{6}\left(3 m^3+84 m^2 n+10 m^2+90 m n^2+6 m n+21 m-6 n^2+66 n+14\right)$  \\ \bs

\bs$b_{5}=$ & $-\frac{m}{6}\left(3 m^3+84 m^2 n+10 m^2+90 m n^2+6 m n+21 m-6 n^2+66 n+14\right)$ \\ \bs

\bs$b_{6}=$ & $-\frac{2n}{3}\left(3 m^2+12 m n-3 m+n^2-3 n+2\right)$ \\ \bs

\bs\multirow{2}{*}{$b_{7}=$} & $\frac{1}{6}\big(-12 m^3+90 m^2 n^2-138 m^2 n-24 m^2+84 m n^3-150 m n^2+66 m n-$ \\ \bs
& $12 m+3 n^4-50 n^3+45 n^2-46 n\big)$ \\ \bs

\bs\multirow{2}{*}{$b_{8}=$} & $\frac{1}{12}\big(3 m^4+84 m^3 n+26 m^3-90 m^2 n^2+66 m^2 n+45 m^2-168 m n^3-$ \\ \bs
& $66 m n^2-66 m n+22 m-6 n^4-36 n^3-66 n^2-36 n\big)$ \\ \bs

\bs$b_{9}=$ & $2 m n \left(3 m^2+9 m n-2 m+3 n^2-2 n+3\right)$ \\ \bs

\bs$b_{10}=$ & $6 m n \left(m^2+3 m n+n^2+1\right)$ \\ \bs

\bs$b_{11}=$ & $-2 m n \left(3 m^2+9 m n-2 m+3 n^2-2 n+3\right)$ \\ \bs

\bs$b_{12}=$ & $-4 m n (m+n)$ \\ \bs

\bs$b_{13}=$ & $-6 m n \left(m^2+3 m n+n^2+1\right)$ \\ \bs

\bs$b_{14}=$ & $m n \left(3 m^2+9 m n+2 m+3 n^2+2 n+3\right)$ \\ \bs

\bs$b_{15}=$ & $-2 m n \left(m^2+3 m n+n^2+1\right)$ \\ \bs

\bs$b_{16}=$ & $-\frac{n}{6}n\left(30 m^2 n-6 m^2+28 m n^2-18 m n+26 m+n^3-6 n^2+11 n-6\right)$ \\ \bs

\bs$b_{17}=$ & $-2 m n \left(m^2+3 m n+n^2+1\right)$ \\ \bs

\bs\multirow{2}{*}{$b_{18}=$} & $\frac{1}{12}\big(m^4+28 m^3 n-2 m^3+90 m^2 n^2-18 m^2 n-m^2+56 m n^3+18 m n^2+$ \\ \bs
& $66 m n+2 m+2 n^4+12 n^3+22 n^2+12 n\big)$ \\ \bs

\bs$b_{19}=$ & $-2 m n \left(m^2+3 m n+n^2+1\right)$ \\ \bs

\bs$b_{20}=$ & $2 m n \left(m^2+3 m n+n^2+1\right)$ \\ \bs

\bs$b_{21}=$ & $m n \left(m^2+3 m n-2 m+n^2-2 n+1\right)$ \\ \bs

\bs$b_{22}=$ & $2 m n \left(m^2+3 m n+n^2+1\right)$ \\ \bs

\bs\multirow{3}{*}{$b_{23}=$} & $\frac{m}{6}\big(3 m^3+84 m^2 n+10 m^2+90 m n^2+6 m n+21 m-6 n^2+66 n+14+$ \\ \bs
& $12n\left(3 m^2+9 m n-2 m+3 n^2-2 n+3\right)\psi_{0}(n)+72n(m^2+3 m n+$ \\ \bs
& $n^2+1)\psi_{0}(n-m)\big)$ \\ \bs

\bs$b_{24}=$ & $-6 m n \left(m^2+3 m n+n^2+1\right)$ \\ \bs

\bs$b_{25}=$ & $-2 m n \left(m^2+3 m n+n^2+1\right)$ \\ \bs

\hline\hline\bs
\caption{Coefficients in~(\ref{eq:IBf}) of $I_{B}$.} \\
\label{t:IBc}
\end{longtable}

\begin{eqnarray}
\fl I_{C}&=&c_{1}+c_{2}\psi_{0}(n)+c_{3}\psi_{0}(n-m)+c_{4}\psi_{0}(1)\psi_{0}(n-m)+c_{5}\psi_{0}(m)\psi_{0}(n-m)+\nonumber\\
\fl&&c_{6}\psi_{0}^{2}(n)+c_{7}\psi_{0}(n)\psi_{0}(n-m)+c_{8}\psi_{0}^{2}(n-m)+c_{9}\psi_{0}(1)\psi_{0}(n)\psi_{0}(n-m)+\nonumber\\
\fl&&c_{10}\psi_{0}(m)\psi_{0}(n)\psi_{0}(n-m)+c_{11}\psi_{0}^{3}(n)+c_{12}\psi_{0}^{2}(n)\psi_{0}(n-m)+\nonumber\\
\fl&&c_{13}\psi_{0}(1)\psi_{0}^{2}(n-m)+c_{14}\psi_{0}(m)\psi_{0}^{2}(n-m)+c_{15}\psi_{0}(n)\psi_{0}^{2}(n-m)+\nonumber\\
\fl&&c_{16}\psi_{0}^{3}(n-m)+c_{17}\psi_{1}(n)+c_{18}\psi_{0}(n)\psi_{1}(n)+c_{19}\psi_{0}(n-m)\psi_{1}(n)+c_{20}\psi_{1}(n-m)+\nonumber\\
\fl&&c_{21}\psi_{0}(1)\psi_{1}(n-m)+c_{22}\psi_{0}(m)\psi_{1}(n-m)+c_{23}\psi_{0}(n)\psi_{1}(n-m)+\nonumber\\
\fl&&c_{24}\psi_{0}(n-m)\psi_{1}(n-m)+c_{25}\psi_{2}(n)+c_{26}\psi_{2}(n-m)+c_{27}\sum_{k=1}^{m}\frac{\psi_{0}(k+n-m)}{k}+\nonumber\\
\fl&&c_{28}\sum_{k=1}^{m}\frac{\psi_{0}^{2}(k+n-m)}{k}+c_{29}\sum_{k=1}^{m}\frac{\psi_{1}(k+n-m)}{k}.\label{eq:ICf}
\end{eqnarray}

\begin{longtable}[h!]{cl}
\caption{Coefficients in~(\ref{eq:ICf}) of $I_{C}$.}\\
\ns\ns\ns
\hline\hline

\bs\multirow{2}{*}{$c_{1}=$} & $\frac{m}{288}\big(37 m^3+4012 m^2 n+1114 m^2+4410 m n^2+2430 m n+1043 m+$ \\ \bs
& $84 n^3+90 n^2-82 n-34\big)$ \\ \bs

\bs\multirow{2}{*}{$c_{2}=$} & $-\frac{1}{24}\big(12 m^3 n+414 m^2 n^2+90 m^2 n-36 m^2+364 m n^3+342 m n^2-$ \\ \bs
& $142 m n+12 m+7 n^4+10 n^3-127 n^2+134 n+12\big)$ \\ \bs

\bs\multirow{2}{*}{$c_{3}=$} & $-\frac{1}{24}\big(7 m^4+352 m^3 n+106 m^3+264 m^2 n+125 m^2-352 m n^3+$ \\ \bs
& $26 m-7 n^4-10 n^3+55 n^2+106 n+36\big)$ \\ \bs

\bs$c_{4}=$ & $\frac{m}{2}\left(m^3+28 m^2 n+2 m^2+30 m n^2-6 m n+5 m-6 n^2+20 n+4\right)$ \\ \bs

\bs$c_{5}=$ & $-\frac{m}{2}\left(m^3+28 m^2 n+2 m^2+30 m n^2-6 m n+5 m-6 n^2+20 n+4\right)$ \\ \bs

\bs$c_{6}=$ & $\frac{1}{2}\left(-6 m^2 n+3 m^2-24 m n^2+15 m n+3 m-2 n^3+6 n^2-4 n\right)$ \\ \bs

\bs\multirow{2}{*}{$c_{7}=$} & $-\frac{1}{2}\big(6 m^3-30 m^2 n^2+60 m^2 n+12 m^2-28 m n^3+48 m n^2-20 m n+$ \\ \bs
& $6 m-n^4+16 n^3-17 n^2+14 n\big)$ \\ \bs

\bs\multirow{2}{*}{$c_{8}=$} & $\frac{1}{4}\big(m^4+28 m^3 n+10 m^3-30 m^2 n^2+30 m^2 n+17 m^2-56 m n^3-$ \\ \bs
& $18 m n^2-20 m n+8 m-2 n^4-12 n^3-22 n^2-12 n\big)$ \\ \bs

\bs$c_{9}=$ & $6 m n \left(m^2+3 m n-m+n^2-n+1\right)$ \\ \bs

\bs$c_{10}=$ & $-6 m n \left(m^2+3 m n-m+n^2-n+1\right)$ \\ \bs

\bs$c_{11}=$ & $mn$ \\ \bs

\bs$c_{12}=$ & $-6 m n (m + n)$ \\ \bs

\bs$c_{13}=$ & $6 m n \left(m^2+3 m n+n^2+1\right)$ \\ \bs

\bs$c_{14}=$ & $-6 m n \left(m^2+3 m n+n^2+1\right)$ \\ \bs

\bs$c_{15}=$ & $3 m n \left(m^2+3 m n+m+n^2+n+1\right)$ \\ \bs

\bs$c_{16}=$ & $-2 m n \left(m^2+3 m n+n^2+1\right)$ \\ \bs

\bs\multirow{2}{*}{$c_{17}=$} & $\frac{1}{4}\big(4 m^3-30 m^2 n^2+30 m^2 n+6 m^2-28 m n^3+30 m n^2-20 m n+$ \\ \bs
& $2 m-n^4+6 n^3-11 n^2+6 n\big)$ \\ \bs

\bs$c_{18}=$ & $3 m n (m+n)$ \\ \bs

\bs$c_{19}=$ & $-3 m n \left(m^2+3 m n+n^2+1\right)$ \\ \bs

\bs\multirow{2}{*}{$c_{20}=$} & $\frac{1}{4}\big(\!-4 m^3+30 m^2 n^2-18 m^2 n-6 m^2+28 m n^3+6 m n^2+20 m n-$ \\ \bs
& $2 m+n^4+6 n^3+11 n^2+6 n\big)$ \\ \bs

\bs$c_{21}=$ & $-3 m n \left(m^2+3 m n+n^2+1\right)$ \\ \bs

\bs$c_{22}=$ & $3 m n \left(m^2+3 m n+n^2+1\right)$ \\ \bs

\bs$c_{23}=$ & $-3 m n (m+n)$ \\ \bs

\bs$c_{24}=$ & $3 m n \left(m^2+3 m n+n^2+1\right)$ \\ \bs

\bs$c_{25}=$ & $\frac{mn}{2}\left(m^2+3 m n+n^2+1\right)$ \\ \bs

\bs$c_{26}=$ & $-\frac{mn}{2}\left(m^2+3 m n+n^2+1\right)$ \\ \bs

\bs\multirow{3}{*}{$c_{27}=$} & $\frac{m}{2}\big(m^3+28 m^2 n+2 m^2+30 m n^2-6 m n+5 m-6 n^2+20 n+4+$ \\ \bs
& $12n\left(m^2+3 m n-m+n^2-n+1\right)\psi_{0}(n)+24n\big(m^2+3 m n+$ \\ \bs
& $n^2+1\big)\psi_{0}(n-m)\big)$ \\ \bs

\bs$c_{28}=$ & $-6 m n \left(m^2+3 m n+n^2+1\right)$ \\ \bs

\bs$c_{29}=$ & $-3 m n \left(m^2+3 m n+n^2+1\right)$ \\ \bs

\hline\hline\bs
\caption{Coefficients in~(\ref{eq:ICf}) of $I_{C}$.}
\label{t:ICc}
\end{longtable}
\noindent Finally, inserting~(\ref{eq:IAf}),~(\ref{eq:IBf}), and~(\ref{eq:ICf}) into~(\ref{eq:ABC}), we observe substantial cancellation among the terms in $I_{A}-3I_{B}+2I_{C}$. In particular, polygamma functions of argument $n-m$ and the three types of unsimplifiable sums
\begin{eqnarray}
&\sum_{k=1}^{m}\frac{\psi_{0}(k+n-m)}{k} \label{eq:ub1} \\
&\sum_{k=1}^{m}\frac{\psi_{0}^{2}(k+n-m)}{k} \label{eq:ub2} \\
&\sum_{k=1}^{m}\frac{\psi_{1}(k+n-m)}{k}. \label{eq:ub3}
\end{eqnarray}
cancel completely. The surviving terms give us
\begin{eqnarray}
\fl I_{A}-3I_{B}+2I_{C}&=&2c_{25}\psi_{2}(n)+2c_{18}\psi_{0}(n)\psi_{1}(n)+\left(2c_{17}-3b_{16}\right)\psi_{1}(n)+2c_{11}\psi_{0}^{3}(n)+\nonumber\\
\fl&&\left(2c_{6}-3b_{6}\right)\psi_{0}^{2}(n)+\left(a_2-3b_2+2c_2\right)\psi_{0}(n)+a_{1}-3b_{1}+2c_{1} \label{eq:T3f}\\
\fl&=&mn\left(m^2+3mn+n^2+1\right)\psi_{2}(n)+6mn(m+n)\psi_{0}(n)\psi_{1}(n)+\nonumber\\
\fl&&m\left(2m^2+12mn+3m+6n^{2}+3n+1\right)\psi_{1}(n)+2mn\psi_{0}^{3}(n)+\nonumber\\
\fl&&3m(m+3n+1)\psi_{0}^{2}(n)+6m(m+n+1)\psi_{0}(n)+m(m+1),
\end{eqnarray}
which is the desired identity~(\ref{eq:T3R}). This completes the proof of the main result~(\ref{eq:k3}).

Before the end of the paper, a few remarks are in order. Firstly, note that the results~(\ref{eq:IAf}),~(\ref{eq:IBf}), and~(\ref{eq:ICf}) become indeterminate when $m=n$ since some of the polygamma functions approach infinity. The resulting identity~(\ref{eq:T3R}) is still valid for $m=n$ due to the cancellation of all polygamma functions of argument $n-m$ as observed in~(\ref{eq:T3f}). On the other hand, the indeterminacy can be also resolved by taking appropriate limits using~(106). Secondly, note that as a result of applying the semi closed-form formulas, we in fact ended up with four types of unsimplifiable summations. Namely, in addition to~(\ref{eq:ub1}),~(\ref{eq:ub2}), and~(\ref{eq:ub3}), we also have
\begin{equation} \label{eq:ub0}
\sum_{k=1}^{m}\frac{\psi_{0}(k)}{k+n-m}.
\end{equation}
The sum~(\ref{eq:ub0}), however, can be expressed by the sum~(\ref{eq:ub1}) through the identity
\begin{eqnarray}
\fl\sum_{k=1}^{m}\left(\frac{\psi_{0}(k+a)}{k}+\frac{\psi_{0}(k)}{k+a}\right)&=&\psi_{0}(m+1)\psi_{0}(a+m+1)-\psi_{0}(1)\psi_{0}(a+1)+\frac{1}{a}\times\nonumber\\
&&\left(\psi_{0}(a+m+1)-\psi_{0}(a+1)-\psi_{0}(m+1)+\psi_{0}(1)\right),
\end{eqnarray}
which is a special case of the result~\cite[Eq.~(23)]{Milgram}
\begin{eqnarray}
\fl\sum_{k=1}^{m}\left(\frac{\psi_{0}(k+a)}{k+b}+\frac{\psi_{0}(k+b)}{k+a}\right)&=&\psi_{0}(a+m+1)\psi_{0}(b+m+1)-\psi_{0}(a+1)\psi_{0}(b+1)+\nonumber\\
&&\frac{1}{a-b}(\psi_{0}(a+m+1)-\psi_{0}(b+m+1)-\psi_{0}(a+1)+\nonumber\\
&&\psi_{0}(b+1)).
\end{eqnarray}
In the limit $a$ approaches $b$, the above identity reduces to~\cite[Eq.~(26)]{Milgram}
\begin{equation}
\fl\sum_{k=1}^{m}\frac{\psi_{0}(k+a)}{k+a}=\frac{1}{2}\left(\psi_{1}(a+m+1)-\psi_{1}(a+1)+\psi_{0}^{2}(a+m+1)-\psi_{0}^{2}(a+1)\right),
\end{equation}
which has also been utilized in simplifying the summations. Since no further relation seems to exist among~(\ref{eq:ub1}),~(\ref{eq:ub2}), and~(\ref{eq:ub3}), we call these sums bases in representing the
unsimplifiable summations in $I_{A}$, $I_{B}$, and $I_{C}$. In the simplification, we also find that the unsimplifiable sum
\begin{equation}\label{eq:vb1}
\sum_{k=1}^{m}\frac{\psi_{0}^{2}(k+a)}{k+a}
\end{equation}
always comes in pairs with the unsimplifiable sum
\begin{equation}\label{eq:vb2}
\sum_{k=1}^{m}\frac{\psi_{1}(k+a)}{k+a}.
\end{equation}
It can be verified by the principle discussed in~\ref{sec:ap1-re} the following closed-form relation of the two unsimplifiable sums
\begin{eqnarray}
\fl\sum_{k=1}^{m}\frac{\psi_{0}^{2}(k+a)+\psi_{1}(k+a)}{k+a}&=&\frac{1}{3}\Big(\psi_{2}(a+m+1)-\psi_{2}(a+1)+3\psi_{0}(a+m+1)\times\nonumber\\
&&\psi_1(a+m+1)-3\psi_{0}(a+1)\psi_{1}(a+1)+\psi_{0}^{3}(a+m+1)-\nonumber\\
&&\psi_{0}^{3}(a+1)\Big),
\end{eqnarray}
which excludes~(\ref{eq:vb1}) or~(\ref{eq:vb2}) being considered as an unsimplifiable sum basis. Finally, as seen from this section, in computing the integrals in~(\ref{eq:IAI}),~(\ref{eq:IBI}), and~(\ref{eq:ICI}) an essential task that the formulation in~\cite{Bianchi19} will also inevitably end up with is to capture the cancellation of each of the unsimplifiable sums~(\ref{eq:ub1}),~(\ref{eq:ub2}),~(\ref{eq:ub3}),~(\ref{eq:ub0}). This in particular requires deriving tailor-made semi closed-form formulas as listed in~\ref{sec:ap1-sc} and~\ref{sec:ap2-sc}, which are unavailable in the computer algebra system Mathematica~\cite{KG}. It is therefore unclear the statement in~\cite{Bianchi19} that the calculations were performed by using Mathematica to yield the result~(\ref{eq:k3}).

\section*{Acknowledgments}
The author thanks Bjordis Collaku and Xhoendi Collaku for helping with the simplification task. The author also wishes to thank Sean O'Rourke and Nicholas Witte for the discussion.

\appendix

\section{Polygamma summation identities of the first type}\label{App1}
In this appendix, we list finite summation identities of polygamma functions of the type
\begin{equation}\label{eq:A1}
\sum_{k=1}^{n}k^{c}\psi_{j_{1}}^{b_{1}}(k+a_{1})\psi_{j_{2}}^{b_{2}}(k+a_{2})\cdots\psi_{j_{m}}^{b_{m}}(k+a_{m})
\end{equation}
(hereinafter referred to as the first type) useful in the simplification process in Section~\ref{sec:simp}, where $\{a_{l}\}_{l=1}^{m}$, $\{b_{l}\}_{l=1}^{m}$, $\{j_{l}\}_{l=1}^{m}$, and $c$, are non-negative integers. We list the corresponding closed-form identities in~\ref{sec:ap1-c} and semi closed-form identities (that contain an unsimplifiable term) in~\ref{sec:ap1-sc}. Some remarks on derivation and implementation of the listed formulas are provided in~\ref{sec:ap1-re}.

\subsection{Close-form expressions}\label{sec:ap1-c}
\begin{equation}\label{eq:A2}
\fl\sum_{k=1}^n\psi_{0}(k+a)=(a+n)\psi_{0}(a+n+1)-a\psi_{0}(a+1)-n
\end{equation}
\begin{eqnarray}\label{eq:A3}
\fl\sum_{k=1}^{n}k\psi_{0}(k+a)&=&\frac{1}{2}\left(-a^2+a+n^2+n\right)\psi_{0}(a+n+1)+\frac{1}{2}(a-1)a\psi_{0}(a+1)+\nonumber \\
&&\frac{1}{4}n(2a-n-3)
\end{eqnarray}
\begin{eqnarray}\label{eq:A4}
\fl\sum_{k=1}^{n}k^{2}\psi_{0}(k+a)&=&\frac{1}{6}\left(2a^3-3a^2+a+2n^3+3n^2+n\right)\psi_{0}(a+n+1)-\frac{1}{6}a\Big(2a^2-\nonumber\\
&&3a+1\Big)\psi_{0}(a+1)-\frac{1}{36}n\left(12a^2-6an-24a+4n^2+15n+17\right)
\end{eqnarray}
\begin{eqnarray}\label{eq:A5}
\fl\sum_{k=1}^{n}k^3\psi_{0}(k+a)&=&\frac{1}{4} \left(a^4-2a^3+a^2-n^4-2n^3-n^2\right)\psi_{0}(a+n+1)+\frac{1}{4}(a-1)^{2}a^{2}\times\nonumber\\
&&\psi_{0}(a+1)-\frac{1}{48}n\Big(\!-12a^3+6a^{2}n+30a^2-4an^2-18an-26a+\nonumber\\
&&3n^3+14n^2+21n+10\Big)
\end{eqnarray}
\begin{eqnarray}\label{eq:A6}
\fl\sum_{k=1}^{n}\psi_{0}^{2}(k+a)&=&(a+n)\psi_{0}^{2}(a+n+1)-(2a+2n+1)\psi_{0}(a+n+1)-a\psi_{0}^{2}(a+1)+\nonumber\\
&&(2a+1)\psi_{0}(a+1)+2n
\end{eqnarray}
\begin{eqnarray}\label{eq:A7}
\fl\sum_{k=1}^{n}k\psi_{0}^{2}(k+a)&=&\frac{1}{2}\left(-a^2+a+n^2+n\right)\psi_{0}^{2}(a+n+1)+\frac{1}{4}\Big(6a^2+4an-2a-2n^2-\nonumber\\
&&6n-2\Big)\times\psi_{0}(a+n+1)+\frac{1}{2}(a-1)a\psi_{0}^{2}(a+1)+\frac{1}{4}\Big(\!-6a^2+2a+\nonumber\\
&&2\Big)\psi_{0}(a+1)+\frac{1}{4}n(-6a+n+5)
\end{eqnarray}
\begin{eqnarray}\label{eq:A8}
\fl\sum_{k=1}^{n}k^2\psi_{0}^{2}(k+a)&=&\frac{1}{6}\left(2a^3-3a^2+a+2n^3+3n^2+n\right)\psi_{0}^{2}(a+n+1)-\frac{1}{18}\Big(22a^3+\nonumber\\
&&12a^{2}n-21a^2-6an^2-24an-a+4n^3+15n^2+17n+3\Big)\times\nonumber\\
&&\psi_{0}(a+n+1)-\frac{1}{6}a\left(2a^2-3a+1\right)\psi_{0}^{2}(a+1)+\frac{1}{18}\Big(22a^3-\nonumber\\
&&21a^2-a+3\Big)\psi_{0}(a+1)+\frac{1}{108}n\Big(132a^2-30an-192a+8n^2+\nonumber\\
&&+39n+79\Big)
\end{eqnarray}
\begin{eqnarray}\label{eq:A9}
\fl\sum_{k=1}^{n}k^3\psi_{0}^{2}(k+a)&=&-\frac{1}{4}\left(a^4-2a^3+a^2-n^4-2n^3-n^2\right)\psi_{0}^{2}(a+n+1)+\frac{1}{24}\Big(25a^4+\nonumber\\
&&12a^{3}n-38a^3-6a^{2}n^2-30a^{2}n+11a^2+4an^3+18an^2+26an+\nonumber\\
&&2a-3n^4-14n^3-21n^2-10n\Big)\psi_{0}(a+n+1)+\frac{1}{4}(a-1)^{2}a^2\times\nonumber\\
&&\psi_{0}^{2}(a+1)-\frac{1}{24}a\left(25a^3-38a^2+11a+2\right)\psi_{0}(a+1)+\frac{1}{288}n\times\nonumber\\
&&\Big(\!-300a^3+78a^{2}n+606a^2-28an^2-162an-410a+9n^3+\nonumber\\
&&50n^2+111n+118\Big)
\end{eqnarray}
\begin{eqnarray}\label{eq:A10}
\fl\sum_{k=1}^{n}\psi_{0}^{3}(k+a)&=&-\frac{1}{2}\psi_{1}(a+n+1)+\frac{1}{2}\psi_{1}(a+1)+(a+n)\psi_{0}^{3}(a+n+1)-\frac{3}{2}(2a+\nonumber\\
&&2n+1)\psi_{0}^{2}(a+n+1)+3(2a+2n+1)\psi_{0}(a+n+1)-a\times\nonumber\\
&&\psi_{0}^{3}(a+1)+\frac{3}{2}(2a+1)\psi_{0}^{2}(a+1)-3(2a+1)\psi_{0}(a+1)-6n
\end{eqnarray}
\begin{eqnarray}\label{eq:A11}
\fl\sum_{k=1}^{n}k\psi_{0}^{3}(k+a)&=&\frac{1}{4}\left(2a-1\right)\psi_{1}(a+n+1)+\frac{1}{4}\left(-2a+1\right)\psi_{1}(a+1)+\frac{1}{2}\Big(\!-a^2+\nonumber\\
&&a+n^2+n\Big)\psi_{0}^{3}(a+n+1)+\frac{3}{4}\left(3a^2+2an-a-n^2-3n-1\right)\times\nonumber\\
&&\psi_{0}^{2}(a+n+1)+\frac{1}{8}\left(-42a^2-36an+6a+6n^2+30n+14\right)\times\nonumber\\
&&\psi_{0}(a+n+1)+\frac{1}{2}(a-1)a\psi_{0}^{3}(a+1)+\frac{3}{4}\left(-3a^2+a+1\right)\times\nonumber\\
&&\psi_{0}^{2}(a+1)+\frac{1}{4}\left(21a^2-3a-7\right)\psi_{0}(a+1)+\frac{1}{8}\Big(42an-3n^2-\nonumber\\
&&27n\Big)
\end{eqnarray}
\begin{eqnarray}\label{eq:A12}
\fl\sum_{k=1}^{n}k^{2}\psi_{0}^{3}(k+a)&=&\frac{1}{12}\left(-6a^2+6a-1\right)\psi_{1}(a+n+1)+\frac{1}{12}\left(6a^2-6a+1\right)\psi_{1}(a+1)-\nonumber\\
&&\frac{1}{6}\left(2a^3-3a^2+a+2n^3+3n^2+n\right)\psi_{0}^{3}(a+n+1)-\frac{1}{12}\Big(22a^3+\nonumber\\
&&12a^{2}n-21a^2-6an^2-24an-a+4n^3+15n^2+17n+3\Big)\times\nonumber\\
&&\psi_{0}^{2}(a+n+1)+\frac{1}{36}\Big(170a^3+132a^{2}n-123a^2-30an^2-192an-\nonumber\\
&&47a+8n^3+39n^2+79n+33\Big)\psi_{0}(a+n+1)-\frac{1}{6}a\left(2a^2-3a+1\right)\times\nonumber\\
&&\psi_{0}^{3}(a+1)+\frac{1}{12}\left(22a^3-21a^2-a+3\right)\psi_{0}^{2}(a+1)-\frac{1}{36}\Big(170a^3-\nonumber\\
&&123a^2-47a+33\Big)\psi_{0}(a+1)+\frac{1}{216}\Big(\!-1020a^{2}n+114an^2+\nonumber\\
&&1248an-16n^3-105n^2-365n\Big)
\end{eqnarray}
\begin{eqnarray}\label{eq:A13}
\fl\sum_{k=1}^{n}k^{3}\psi_{0}^{3}(k+a)&=&\frac{1}{4}a\left(2a^2-3a+1\right)\psi_{1}(a+n+1)+\frac{1}{4}a\left(-2a^2+3a-1\right)\psi_{1}(a+1)-\nonumber\\
&&\frac{1}{4}\left(a^4-2a^3+a^2-n^4-2n^3-n^2\right)\psi_{0}^{3}(a+n+1)+\frac{1}{16}\Big(25a^4+\nonumber\\
&&12a^{3}n-38a^3-6a^{2}n^2-30a^{2}n+11a^{2}+4an^{3}+18an^2+26an+\nonumber\\
&&2a-3n^4-14n^3-21n^2-10n\Big)\psi_{0}^{2}(a+n+1)-\frac{1}{96}\Big(415a^4+\nonumber\\
&&300a^{3}n-530a^3-78a^{2}n^{2}-606a^{2}n+17a^2+28an^3+162an^2+\nonumber\\
&&410an+146a-9n^4-50n^3-111n^2-118n-36\Big)\psi_{0}(a+n+1)+\nonumber\\
&&\frac{1}{4}(a-1)^{2}a^{2}\psi_{0}^{3}(a+1)-\frac{1}{16}a\left(25a^3-38a^2+11a+2\right)\psi_{0}^{2}(a+1)+\nonumber\\
&&\frac{1}{96}\left(415a^4-530a^3+17a^2+146a-36\right)\psi_{0}(a+1)+\frac{1}{1152}\times\nonumber\\
&&\Big(4980a^{3}n-690a^{2}n^{2}-8850a^{2}n+148an^3+1134an^2+4790an-\nonumber\\
&&27n^4-182n^3-525n^2-850n\Big)
\end{eqnarray}
\begin{equation}\label{eq:A14}
\fl\sum_{k=1}^{n}\psi_{1}(k+a)=(a+n)\psi_{1}(a+n+1)-a\psi_{1}(a+1)+\psi_{0}(a+n+1)-\psi_{0}(a+1)
\end{equation}
\begin{eqnarray}\label{eq:A15}
\fl\sum_{k=1}^{n}k\psi_{1}(k+a)&=&\frac{1}{2}\Big(\!\left(-a^2+a+n^2+n\right)\psi_{1}(a+n+1)+(a-1)a\psi_{1}(a+1)+\nonumber\\
&&(-2a+1)\psi_{0}(a+n+1)+(2a-1)\psi_{0}(a+1)+n\Big)
\end{eqnarray}
\begin{eqnarray}\label{eq:A16}
\fl\sum_{k=1}^{n}k^{2}\psi_{1}(k+a)&=&\frac{1}{6}\Big(\!\left(2a^3-3a^2+a+2n^3+3n^2+n\right)\psi_{1}(a+n+1)+a(a-1)\times\nonumber\\
&&(-2a+1)\psi_{1}(a+1)+\left(6a^2-6a+1\right)\psi_{0}(a+n+1)+\Big(\!-6a^2+\nonumber\\
&&6a-1\Big)\psi_{0}(a+1)-4an+n^2+4n\Big)
\end{eqnarray}
\begin{eqnarray}\label{eq:A17}
\fl\sum_{k=1}^{n}k^{3}\psi_{1}(k+a)&=&\frac{1}{24}\Big(6\left(-a^4+2a^3-a^2+n^4+2n^3+n^2\right)\psi_{1}(a+n+1)+6\Big(a^4-\nonumber\\
&&2a^3+a^2\Big)\psi_{1}(a+1)-12a\left(2a^2-3a+1\right)\psi_{0}(a+n+1)+12a\times\nonumber\\
&&\left(2a^2-3a+1\right)\psi_{0}(a+1)+18a^{2}n-6an^2-30an+2n^3+9n^2+\nonumber\\
&&13n\Big)
\end{eqnarray}
\begin{eqnarray}\label{eq:A18}
\fl\sum_{k=1}^{n}\psi_{2}(k+a)&=&(a+n)\psi_{2}(a+n+1)-a\psi_{2}(a+1)+2\psi_{1}(a+n+1)-\nonumber\\
&&2\psi_{1}(a+1)
\end{eqnarray}
\begin{eqnarray}\label{eq:A19}
\fl\sum_{k=1}^{n}k\psi_{2}(k+a)&=&\frac{1}{2}\left(-a^2+a+n^2+n\right)\psi_{2}(a+n+1)+\frac{1}{2}a(a-1)\psi_{2}(a+1)+\nonumber\\
&&(-2a+1)\psi_{1}(a+n+1)+(2a-1)\psi_{1}(a+1)-\psi_{0}(a+n+1)+\nonumber\\
&&\psi_{0}(a+1)
\end{eqnarray}
\begin{eqnarray}\label{eq:A20}
\fl\sum_{k=1}^{n}k^{2}\psi_{2}(k+a)&=&\frac{1}{6}\Big(\!\left(2a^3-3a^2+a+2n^3+3n^2+n\right)\psi_{2}(a+n+1)+a\Big(\!-2a^2+\nonumber\\
&&3a-1\Big)\psi_{2}(a+1)+2\left(6a^2-6a+1\right)\psi_{1}(a+n+1)+2\Big(\!-6a^2+\nonumber\\
&&6a-1\Big)\psi_{1}(a+1)+6(2a-1)\psi_{0}(a+n+1)+6(-2a+1)\times\nonumber\\
&&\psi_{0}(a+1)-4n\Big)
\end{eqnarray}
\begin{eqnarray}\label{eq:A21}
\fl\sum_{k=1}^{n}k^{3}\psi_{2}(k+a)&=&\frac{1}{4}\Big(\!\left(-a^4+2a^3-a^2+n^4+2n^3+n^2\right)\psi_{2}(a+n+1)+(a-1)^{2}a^{2}\times\nonumber\\
&&\psi_{2}(a+1)+4a\left(-2a^2+3a-1\right)\psi_{1}(a+n+1)+4a\left(2a^2-3a+1\right)\times\nonumber\\
&&\psi_{1}(a+1)-2\left(6a^2-6a+1\right)\psi_{0}(a+n+1)+2\left(6a^2-6a+1\right)\times\nonumber\\
&&\psi_{0}(a+1)+6an-n^2-5n\Big)
\end{eqnarray}
\begin{eqnarray}\label{eq:A22}
\fl\sum_{k=1}^{n}\psi_{0}(k+a)\psi_{1}(k+a)&=&(a+n)\psi_{0}(a+n+1)\psi_{1}(a+n+1)-a\psi_{0}(a+1)\psi_{1}(a+1)-\nonumber\\
&&\frac{1}{2}(2a+2n+1)\psi_{1}(a+n+1)+\frac{1}{2}(2a+1)\psi_{1}(a+1)+\nonumber\\
&&\frac{1}{2}\psi_{0}^{2}(a+n+1)-\psi_{0}(a+n+1)-\frac{1}{2}\psi_{0}^{2}(a+1)+\nonumber\\
&&\psi_{0}(a+1)
\end{eqnarray}
\begin{eqnarray}\label{eq:A23}
\fl\sum_{k=1}^{n}k\psi_{0}(k+a)\psi_{1}(k+a)&=&\frac{1}{4}\Big(2\left(-a^2+a+n^2+n\right)\psi_{0}(a+n+1)\psi_{1}(a+n+1)+\nonumber\\
&&2(a-1)a\psi_{0}(a+1)\psi_{1}(a+1)-\Big(\!-3a^2-2an+a+n^2+\nonumber\\
&&3n+1\Big)\psi_{1}(a+n+1)+\left(-3a^2+a+1\right)\psi_{1}(a+1)+\nonumber\\
&&(1-2a)\psi_{0}^{2}(a+n+1)+(6a+2n-1)\psi_{0}(a+n+1)+\nonumber\\
&&(2a-1)\psi_{0}^{2}(a+1)+(1-6a)\psi_{0}(a+1)-3n\Big)
\end{eqnarray}
\begin{eqnarray}\label{eq:A24}
\fl\sum_{k=1}^{n}k^{2}\psi_{0}(k+a)\psi_{1}(k+a)&=&\frac{1}{36}\Big(6\left(2a^3-3a^2+a+2n^3+3n^2+n\right)\psi_{0}(a+n+1)\times\nonumber\\
&&\psi_1(a+n+1)-6a\left(2a^2-3a+1\right)\psi_{0}(a+1)\psi_{1}(a+1)+\nonumber\\
&&\Big(\!-22a^3-12a^{2}n+21a^2+6an^2+24an+a-4n^3-\nonumber\\
&&15n^2-17n-3\Big)\psi_{1}(a+n+1)+\left(22a^3-21a^2-a+3\right)\times\nonumber\\
&&\psi_{1}(a+1)+3\left(6a^2-6a+1\right)\psi_{0}^{2}(a+n+1)+\Big(\!-66a^2-\nonumber\\
&&24an+42a+6n^2+24n+1\Big)\psi_{0}(a+n+1)-3\Big(6a^2-6a+\nonumber\\
&&1\Big)\psi_{0}^{2}(a+1)-\left(-66a^2+42a+1\right)\psi_{0}(a+1)+44an-\nonumber\\
&&5n^2-32n\Big)
\end{eqnarray}
\begin{eqnarray}\label{eq:A25}
\fl\sum_{k=1}^{n}k^{3}\psi_{0}(k+a)\psi_{1}(k+a)&=&\frac{1}{288}\Big(\!-72\left(a^4-2a^3+a^2-n^4-2n^3-n^2\right)\psi_{0}(a+n+1)\times\nonumber\\
&&\psi_{1}(a+n+1)+72(a-1)^{2}a^2\psi_{0}(a+1)\psi_{1}(a+1)+\nonumber\\
&&\Big(150a^4+72a^{3}n-228a^3-36a^{2}n^{2}-180a^{2}n+66a^2+\nonumber\\
&&24an^3+108an^2+156an+12a-18n^4-84n^3-126n^2-\nonumber\\
&&60n\Big)\psi_{1}(a+n+1)+\left(-150a^4+228a^3-66a^2-12a\right)\times\nonumber\\
&&\psi_{1}(a+1)-72a\left(2a^2-3a+1\right)\psi_{0}^{2}(a+n+1)-12\Big(\!-\nonumber\\
&&50a^3-18a^{2}n+57a^2+6an^2+30an-11a-2n^3-9n^2-\nonumber\\
&&13n-1\Big)\psi_{0}(a+n+1)+72a\left(2a^2-3a+1\right)\psi_{0}^{2}(a+1)+\nonumber\\
&&12\left(-50a^3+57a^2-11a-1\right)\psi_{0}(a+1)-450a^{2}n+\nonumber\\
&&78an^2+606an-14n^3-81n^2-205n\Big)
\end{eqnarray}

\subsection{Semi closed-form expressions}\label{sec:ap1-sc}
\begin{eqnarray}\label{eq:A26}
\fl\sum_{k=1}^{n}\psi_{0}(k+a)\psi_{0}(k)&=&a\sum_{k=1}^{n}\frac{\psi_{0}(k)}{k+a}+n\psi_{0}(a+n+1)\psi_{0}(n+1)-(a+n+1)\psi_{0}(a+n\nonumber\\
&&+1)-n\psi_{0}(n+1)+(a+1)\psi_{0}(a+1)+2n
\end{eqnarray}
\begin{eqnarray}\label{eq:A27}
\fl\sum_{k=1}^{n}k\psi_{0}(k+a)\psi_{0}(k)&=&-\frac{a(a-1)}{2}\sum_{k=1}^{n}\frac{\psi_{0}(k)}{k+a}+\frac{1}{4}\Big(2n(n+1)\psi_{0}(a+n+1)\psi_{0}(n+1)+\nonumber\\
&&\left(a^2-a-n^2-3n-2\right)\psi_{0}(a+n+1)+(2a-n-3)n\nonumber\\
&&\psi_{0}(n+1)-(a-2)(a+1)\psi_{0}(a+1)-3an+n^2+5n\Big)
\end{eqnarray}
\begin{eqnarray}\label{eq:A28}
\fl\sum_{k=1}^{n}k^{2}\psi_{0}(k+a)\psi_{0}(k)&=&\frac{1}{6}a(a-1)(2a-1)\sum_{k=1}^{n}\frac{\psi_{0}(k)}{k+a}+\frac{1}{108}\Big(18n(n+1)(2n+1)\times\nonumber\\
&&\psi_{0}(a+n+1)\psi_{0}(n+1)-3\Big(4a^3-3a^2-a+4n^3+15n^2+\nonumber\\
&&17n+6\Big)\psi_{0}(a+n+1)+3n\Big(\!-12a^2+6an+24a-4n^2-\nonumber\\
&&15n-17\Big)\psi_{0}(n+1)+3\left(4a^3-3a^2-a+6\right)\psi_{0}(a+1)+\nonumber\\
&&n\left(48a^2-15an-96a+8n^2+39n+79\right)\!\Big)
\end{eqnarray}
\begin{eqnarray}\label{eq:A29}
\fl\sum_{k=1}^{n}k^{3}\psi_{0}(k+a)\psi_{0}(k)&=&-\frac{1}{4}a^{2}(a-1)^{2}\sum_{k=1}^{n}\frac{\psi_{0}(k)}{k+a}+\frac{1}{288}\Big(72n^{2}(n+1)^{2}\psi_{0}(a+n+1)\times\nonumber\\
&&\psi_{0}(n+1)+6\Big(3a^4-2a^3-3a^2+2a-3n^4-14n^3-21n^2-\nonumber\\
&&10n\Big)\psi_{0}(a+n+1)+6n\Big(12a^3-6a^{2}n-30a^2+4an^2+18an+\nonumber\\
&&26a-3n^3-14n^2-21n-10\Big)\psi_{0}(n+1)-6a(a-1)(a+1)\times\nonumber\\
&&(3a-2)\psi_{0}(a+1)+n\Big(\!-90a^3+27a^{2}n+219a^2-14an^2-\nonumber\\
&&81an-205a+9n^3+50n^2+111n+118\Big)\Big)
\end{eqnarray}

\subsection{Remarks on the first type summation}\label{sec:ap1-re}
The principles of evaluating finite sums of the first type~(\ref{eq:A1}), that led to the above listed formulas, are simple. The idea is to change the order of sums by first replacing one polygamma function at a time by the definition~(\ref{eq:pg}) and make use of the obtained lower order summation formulas in a recursive manner. We demonstrate the principles by considering the sum below as an example
\begin{equation}\label{A1:eg}
\sum_{k=1}^{n}k^{c}\psi_{0}^{b}(k+a),
\end{equation}
which is a special case of~(\ref{eq:A1}). We first show the recursion in parameter $b$ by the example $b=2$,
\begin{eqnarray}
\fl\sum_{k=1}^{n}\psi_{0}^{2}(k+a)&=&\sum_{k=1}^{n}\sum_{j=1}^{k+a-1}\frac{\psi_{0}(k+a)}{j}-\gamma\sum_{k=1}^{n}\psi_{0}(k+a) \\
&=&\sum_{k=1}^{n}\sum_{j=a+1}^{k+a-1}\frac{\psi_{0}(k+a)}{j}+\left(\sum_{j=1}^{a}\frac{1}{j}-\gamma\right)\sum_{k=1}^{n}\psi_{0}(k+a) \\ &=&\sum_{k=1}^{n}\sum_{j=1}^{k-1}\frac{\psi_{0}(k+a)}{j+a}+\psi_{0}(a+1)\sum_{k=1}^{n}\psi_{0}(k+a) \\
&=&\sum_{j=1}^{n-1}\frac{1}{j+a}\sum_{k=j+1}^{n}\psi_{0}(k+a)+\psi_{0}(a+1)\sum_{k=1}^{n}\psi_{0}(k+a) \\
&=&\sum_{j=1}^{n-1}\frac{1}{j+a}\left(\sum_{k=1}^{n}\psi_{0}(k+a)-\sum_{k=1}^{j}\psi_{0}(k+a)\right)+\psi_{0}(a+1)\sum_{k=1}^{n}\psi_{0}(k+a), \nonumber
\end{eqnarray}
which reduces to evaluating the sum~(\ref{A1:eg}) for $b=1$. To illustrate the recursion in parameter $c$, we consider the example $c=1$ in~(\ref{A1:eg})
\begin{eqnarray}
\sum_{k=1}^{n}k\psi_{0}^{2}(k+a)&=&\sum_{k=1}^{n}\sum_{j=1}^{k}\psi_{0}^{2}(k+a) \\
&=&\sum_{j=1}^{n}\sum_{k=j}^{n}\psi_{0}^{2}(k+a) \\
&=&\sum_{j=1}^{n}\left(\sum_{k=1}^{n}\psi_{0}^{2}(k+a)-\sum_{k=1}^{j-1}\psi_{0}^{2}(k+a)\right),
\end{eqnarray}
which reduces to evaluating the sum~(\ref{A1:eg}) for $c=0$.

Using the principles as shown in the above examples, the listed formulas in this appendix can be derived, which are in fact valid for any non-negative real number $a$. Some of these formulas can be found in the literature. In particular, the formulas~(\ref{eq:A2})--(\ref{eq:A6}),~(\ref{eq:A10}),~(\ref{eq:A22}) are available in~\cite[Chap.~5.1]{Brychkov08}. By keeping in mind the relation between harmonic numbers and polygamma functions~(\ref{eq:pg}), the formulas~(\ref{eq:A2})--(\ref{eq:A13}) and (\ref{eq:A14})--(\ref{eq:A21}) may be also derived from the result~\cite[Th.~2.2]{Spiess90} and the result~\cite[Th.~2.1]{Spiess90}, respectively.

Note that the results~(\ref{eq:A22})--(\ref{eq:A25}) can be also obtained by the relation
\begin{equation}
\frac{\partial}{\partial a}\sum_{k=1}^{n}k^{c}\psi_{0}^{2}(k+a)=2\sum_{k=1}^{n}k^{c}\psi_{0}(k+a)\psi_{1}(k+a),
\end{equation}
and that the semi closed-form expressions~(\ref{eq:A26})--(\ref{eq:A29}) reduce to the corresponding closed-form ones~(\ref{eq:A6})--(\ref{eq:A9}) when $a=0$.

We also point out that currently the computer algebra system Mathematica is only able to evaluate into closed-form expressions the sum of the first type~(\ref{eq:A1}) when $b_{1}=1$, $b_{2}=\cdots=b_{m}=0$, i.e., the sum~\cite{KG}
\begin{equation}
\sum_{k=1}^{n}k^{c}\psi_{j}(k+a).
\end{equation}
We are working with Wolfram Research to implement the polygamma summation~(\ref{eq:A1}) in an algorithmic manner into future versions of Mathematica.

\section{Polygamma summation identities of the second type}\label{App2}
In this appendix, we list finite summation identities of the type
\begin{equation}\label{eq:B1}
S_{f}(m,n)=\sum_{k=1}^{m}\frac{(n-k)!}{(m-k)!}f(k),~~~~~~m\leq n,
\end{equation}
(hereinafter referred to as the second type) that are utilized in the simplification process in Section~\ref{sec:simp}. Here, $f(k)$ is referred to as the test function that may involve a polygamma function and $(n-k)!/(m-k)!$ is referred to as the summation kernel.

We list the corresponding closed-form and semi closed-form identities in~\ref{sec:ap2-c} and~\ref{sec:ap2-sc}, respectively. The strategy in deriving the listed formulas is discussed in~\ref{sec:ap2-re}.

\subsection{Close-form expressions}\label{sec:ap2-c}
\begin{equation}\label{eq:B2}
\fl\sum_{k=1}^{m}\frac{(n-k)!}{(m-k)!}=\frac{n!}{(m-1)!}\frac{1}{n-m+1}
\end{equation}
\begin{equation}\label{eq:B3}
\fl\sum_{k=1}^{m}\frac{(n-k)!}{(m-k)!}\frac{1}{k}=\frac{n!}{m!}\left(\psi_{0}\left(n+1\right)-\psi_{0}\left(n-m+1\right)\right)
\end{equation}
\begin{eqnarray}\label{eq:B4}
\fl\sum_{k=1}^{m}\frac{(n-k)!}{(m-k)!}\psi_{0}(k)&=&\frac{n!}{(m-1)!(n-m+1)}\Bigg(\psi_{0}(n+1)-\psi_{0}(n-m+1)+\psi_{0}(1)-\nonumber\\
&&\frac{1}{n-m+1}\Bigg)
\end{eqnarray}
\begin{eqnarray}\label{eq:B5}
\fl\sum_{k=1}^{m}\frac{(n-k)!}{(m-k)!}\frac{\psi_{0}(k)}{k}&=&\frac{n!}{m!}\Bigg(\frac{1}{2}\Big(\psi_{1}(n+1)-\psi_{1}(n-m+1)+\psi_{0}^{2}(n+1)+\nonumber\\
&&\psi_{0}^{2}(n-m+1)\Big)+\psi_{0}(1)\left(\psi_{0}(n+1)-\psi_{0}(n-m+1)\right)-\nonumber\\
&&\psi_{0}(n+1)\psi_{0}(n-m+1)\Bigg)
\end{eqnarray}
\begin{eqnarray}\label{eq:B6}
\fl\sum_{k=1}^{m}\frac{(n-k)!}{(m-k)!}\frac{\psi_{0}(n+1-k)}{k}&=&\frac{n!}{m!}(\psi_{1}(n+1)-\psi_{1}(n-m+1)+\psi_{0}(n+1)(\psi_{0}(n+1)-\nonumber\\
&&\psi_{0}(n-m+1)))
\end{eqnarray}

\subsection{Semi closed-form expressions}\label{sec:ap2-sc}
\begin{eqnarray}\label{eq:B7}
\fl\sum_{k=1}^{m}\frac{(n-k)!}{(m-k)!}\frac{1}{k+a}&=&\frac{(a+n)!}{(a+m)!}\sum_{k=1}^{m}\frac{(k+n-m-1)!(k+a-1)!}{(k-1)!(k+a+n-m)!}
\end{eqnarray}

\begin{eqnarray}\label{eq:B8}
\fl\sum_{k=1}^{m}\frac{(n-k)!}{(m-k)!}\frac{1}{k^{2}}&=&\frac{n!}{m!}\sum_{k=1}^{m}\frac{\psi_{0}(k+n-m)}{k}+\frac{n!}{m!}\Bigg(\frac{1}{2}\Big(\psi_{1}(n-m+1)-\psi_{1}(n+1)+\nonumber\\
&&\psi_{0}^{2}(n-m+1)-\psi_{0}^{2}(n+1)\Big)+\psi_{0}(n-m)(\psi_{0}(n+1)-\psi_{0}(m+1)-\nonumber\\
&&\psi_{0}(n-m+1)+\psi_{0}(1))\Bigg)
\end{eqnarray}

\begin{eqnarray}\label{eq:B9}
\fl\sum_{k=1}^{m}\frac{(n-k)!}{(m-k)!}\frac{\psi_{0}(k)}{k^{2}}&=&\frac{n!}{2m!}\sum_{k=1}^{m}\frac{\psi_{1}(k+n-m)+\psi_{0}^{2}(k+n-m)}{k}-\frac{n!}{m!}\Big(\psi_{0}(n-m)-\nonumber\\
&&\psi_{0}(1)\Big)\sum_{k=1}^{m}\frac{\psi_{0}(k+n-m)}{k}+\frac{n!}{2m!}\Bigg(\!-\frac{1}{3}\bigg(\psi_{2}(n+1)-\nonumber\\
&&\psi_{2}(n-m+1)+\psi_{0}^{3}(n+1)-\psi_{0}^{3}(n-m+1)+3\psi_{0}(n+1)\times\nonumber\\
&&\psi_{1}(n+1)-3\psi_{0}(n-m+1)\psi_{1}(n-m+1)\bigg)+(\psi_{0}(n-m)-\nonumber\\
&&\psi_{0}(1))\Big(\psi_{1}(n+1)-\psi_{1}(n-m+1)+\psi_{0}^{2}(n+1)-\nonumber\\
&&\psi_{0}^{2}(n-m+1)\Big)-\Big(\psi_{1}(n-m)-\psi_{0}^{2}(n-m)+2\psi_{0}(1)\times\nonumber\\
&&\psi_{0}(n-m)\Big)\Big(\psi_{0}(m+1)-\psi_{0}(n+1)+\psi_{0}(n-m+1)-\nonumber\\
&&\psi_{0}(1)\Big)\Bigg)
\end{eqnarray}

\begin{eqnarray}\label{eq:B10}
\fl\sum_{k=1}^{m}\frac{(n-k)!}{(m-k)!}\frac{\psi_{0}(n+1-k)}{k^{2}}&=&\frac{n!}{m!}\sum_{k=1}^{m}\frac{\psi_{1}(k+n-m)+\psi_{0}(n+1)\psi_{0}(k+n-m)}{k}+\frac{n!}{m!}\times\nonumber\\
&&\Bigg(\frac{1}{2}\psi_{2}(n-m+1)-\frac{1}{2}\psi_{2}(n+1)+\psi_{0}(n-m+1)\times\nonumber\\
&&\psi_{1}(n-m+1)+\psi_{0}(n+1)\Bigg(\frac{1}{2}\Big(\psi_{1}(n-m+1)-\nonumber\\
&&\psi_{1}(n+1)+\psi_{0}^{2}(n-m+1)-\psi_{0}^{2}(n+1)\Big)+\psi_{0}(n-m)\times\nonumber\\
&&(\psi_{0}(n+1)-\psi_{0}(n-m+1)-\psi_{0}(m+1)+\psi_{0}(1))+\nonumber\\
&&\psi_{1}(n-m)-\psi_{1}(n+1)\Bigg)-\psi_{1}(n-m)(\psi_{0}(n-m+1)+\nonumber\\
&&\psi_{0}(m+1)-\psi_{0}(1))+\psi_{0}(n-m)(\psi_{1}(n+1)-\nonumber\\
&&\psi_{1}(n-m+1))\Bigg)
\end{eqnarray}

\begin{eqnarray}\label{eq:B11}
\fl\sum_{k=1}^{m}\frac{(n-k)!}{(m-k)!}\psi_{1}(k)&=&-\frac{n!}{(m-1)!(n-m+1)}\sum_{k=1}^{m}\frac{\psi_{0}(k+n-m)}{k}-\nonumber\\
&&\frac{n!}{(m-1)!(n-m+1)}\Bigg(\frac{1}{2}\Big(\psi_{1}(n-m+1)-\psi_{1}(n)-\psi_{0}^{2}(n)+\nonumber\\
&&\psi_{0}^{2}(n-m+1)\Big)+\psi_{0}(n-m)(\psi_{0}(n)-\psi_{0}(m)-\psi_{0}(n-m+1)+\nonumber\\
&&\psi_{0}(1))-\psi_{1}(1)-\frac{\psi_{0}(n)-\psi_{0}(n-m+1)}{n}-\frac{\psi_{0}(n)}{m}\Bigg)
\end{eqnarray}

\subsection{Remarks on the second type summation}\label{sec:ap2-re}
The generic approach in deriving the summation identities of the second type~(\ref{eq:B1}) relies on finding the recurrence relation between $S_{f}(m,n)$ and $S_{f}(m-1,n-1)$, where the summation terminates after $m$ recursions since $S_{f}(0,n-m)=0$. For a given test function $f(k)$, the recurrence relation can often be found by first rewriting~(\ref{eq:B1}) as
\begin{equation}
S_{f}(m,n)=\sum_{k=1}^{m}\frac{(n-1-k)!}{(m-1-k)!}\frac{n-k}{m-k}f(k),
\end{equation}
where the term
\begin{equation}
\frac{(n-1-k)!}{(m-1-k)!}
\end{equation}
is understood as the new kernel of the sum $S_{f}(m-1,n-1)$ associated with the new test function
\begin{equation}\label{eq:mtf}
\frac{n-k}{m-k}f(k).
\end{equation}
The relation between $S_{f}(m,n)$ and $S_{f}(m-1,n-1)$ can then be obtained by partial fraction decomposition in the variable $k$ of this modified test function~(\ref{eq:mtf}).

To illustrate the above approach, we show in details the derivation of some of the listed formulas as examples. The first example is when $f(k)=1/k$, where the modified test function is decomposed as
\begin{equation}\label{eq:Beg1}
\frac{n-k}{m-k}\frac{1}{k}=\frac{n}{m}\frac{1}{k}+\frac{n-m}{m}\frac{1}{m-k}.
\end{equation}
The corresponding recurrence relation is deduced as
\begin{eqnarray}
S_{f}(m,n)&=&\frac{n}{m}S_{f}(m-1,n-1)+\frac{n-m}{m}\sum_{k=1}^{m}\frac{(n-1-k)!}{(m-k)!} \\
&=&\frac{n}{m}S_{f}(m-1,n-1)+\frac{(n-1)!}{m!},
\end{eqnarray}
where we have used the formula~(\ref{eq:B2}). Iterating $m$ times the above relation leads to the desired expression~(\ref{eq:B3}). The next example is the case $f(k)=\psi_{0}(k)/k$, where similarly as in~(\ref{eq:Beg1}) the test function is decomposed as
\begin{equation}
\frac{n-k}{m-k}\frac{\psi_{0}(k)}{k}=\frac{n}{m}\frac{\psi_{0}(k)}{k}+\frac{n-m}{m}\frac{\psi_{0}(k)}{m-k}.
\end{equation}
The recurrence relation is then calculated as
\begin{eqnarray}
\fl~\!S_{f}(m,n)&\!\!\!\!\!\!\!\!\!\!\!\!\!=&\frac{n}{m}S_{f}(m-1,n-1)+\frac{n-m}{m}\sum_{k=1}^{m}\frac{(n-1-k)!}{(m-k)!}\psi_{0}(k)  \\
&\!\!\!\!\!\!\!\!\!\!\!\!\!=&\frac{n}{m}S_{f}(m-1,n-1)+\frac{(n-1)!}{m!}\left(\psi_{0}(n)-\psi_{0}(n-m+1)+\psi_{0}(1)\right),
\end{eqnarray}
where we have made use of the formula~(\ref{eq:B4}). After $m$ recursions of the above relation, one obtains the claimed identity~(\ref{eq:B5}). We also consider the case $f(k)=1/(k+a)$ as an example, where the modified test function is decomposed as
\begin{equation}
\frac{n-k}{m-k}\frac{1}{k+a}=\frac{n+a}{m+a}\frac{1}{k+a}+\frac{n-m}{m+a}\frac{1}{m-k}.
\end{equation}
The resulting recurrence relation is
\begin{eqnarray}
S_{f}(m,n)&=&\frac{n+a}{m+a}S_{f}(m-1,n-1)+\frac{n-m}{m+a}\sum_{k=1}^{m}\frac{(n-1-k)!}{(m-k)!}\\
&=&\frac{n+a}{m+a}S_{f}(m-1,n-1)+\frac{(n-1)!}{(m-1)!(m+a)},\label{eq:ka}
\end{eqnarray}
where we have utilized the result~(\ref{eq:B2}). The claimed identity~(\ref{eq:B7}) is established after $m$ iterations of the recurrence relation~(\ref{eq:ka}). Note that the formula~(\ref{eq:B3}) is a special case of the formula~(\ref{eq:B7}), which is a useful identity that could transform a summation into one of the listed sums in the appendices. The last example is when $f(k)=1/k^{2}$, where partial fraction decomposition of the modified test function gives
\begin{equation}
\frac{n-k}{m-k}\frac{1}{k^{2}}=\frac{n}{m}\frac{1}{k^{2}}+\frac{n-m}{m^{2}}\frac{1}{k}+\frac{n-m}{m^{2}}\frac{1}{m-k}.
\end{equation}
The recurrence relation is then obtained as
\begin{eqnarray} \fl~\!S_{f}(m,n)&\!\!\!\!\!\!\!\!\!\!\!\!\!=&\frac{n}{m}S_{f}(m-1,n-1)+\frac{n-m}{m^{2}}\left(\sum_{k=1}^{m}\frac{(n-1-k)!}{(m-1-k)!}\frac{1}{k}+\sum_{k=1}^{m}\frac{(n-1-k)!}{(m-k)!}\right)\nonumber\\
&\!\!\!\!\!\!\!\!\!\!\!\!\!=&\frac{n}{m}S_{f}(m-1,n-1)+\frac{(n-1)!(n-m)}{m!m}(\psi_{0}(n)-\psi_{0}(n-m)),
\end{eqnarray}
where we have used the identities~(\ref{eq:B2}) and~(\ref{eq:B3}). Iterating $m$ times the above relation gives the desired result~(\ref{eq:B8}).

Some of formulas in~\ref{App2} exist in the literature: The formula~(\ref{eq:B2}) is the well-known Chu-Vandermonde identity~\cite[p.~99]{Luke} and the formula~(\ref{eq:B3}) can be also obtained via the connection to a hypergeometric function of unit argument as~\cite[p.~111]{Luke}
\begin{eqnarray}
\sum_{k=1}^{m}\frac{(n-k)!}{(m-k)!}\frac{1}{k}&=&\frac{(n-1)!}{(m-1)!}~\!_{3}F_{2}\left(1,1,1-m;2,1-n;1\right)\\
&=&\frac{n!}{m!}\left(\psi_{0}(n+1)-\psi_{0}(n-m+1)\right).
\end{eqnarray}
Moreover, the identity~(\ref{eq:B8}) recently appears in~\cite[Eq.~(A12)]{Wei17}.

Finally, we note that the listed formulas in~\ref{App2} are in fact valid for any positive real number $n$ greater than the integer $m$. This fact provides an alternative derivation of the formulas~(\ref{eq:B6}) and~(\ref{eq:B10}) via the derivative with respect to $n$ on~(\ref{eq:B3}) and~(\ref{eq:B8}), respectively, i.e.,
\begin{equation}
\frac{\partial}{\partial n}\sum_{k=1}^{m}\frac{(n-k)!}{(m-k)!}\frac{1}{k^{c}}=\sum_{k=1}^{m}\frac{(n-k)!}{(m-k)!}\frac{\psi_{0}(n+1-k)}{k^{c}},~~~~~~c=1,2.
\end{equation}

\end{document}